\documentclass[a4paper,USenglish,cleveref,thm-restate]{lipics-v2021}

\title{CrudiTEE: A Stick-and-Carrot Approach to Building Trustworthy Cryptocurrency Wallets with TEEs}

\usepackage[utf8]{inputenc}
\usepackage{xcolor}
\usepackage{amsthm}
\usepackage{amsmath}
\usepackage{amsfonts}
\usepackage{hyperref}
\usepackage{cleveref}
\usepackage{graphicx}
\usepackage{algorithm}
\usepackage[noend]{algpseudocode}
\usepackage{tcolorbox}
\newcommand{\ceil}[1]{\left\lceil #1 \right\rceil}

\usepackage[export]{adjustbox}

\usepackage{enumerate}
\usepackage{dtrt}
\usepackage{xspace}
\usepackage[scaled=0.88]{helvet}

\author{Lulu Zhou}{Yale University, USA}{lulu.zhou@yale.edu}{https://orcid.org/0000-0001-9270-9524}{}
\author{Zeyu Liu}{Yale University, USA}{zeyu.liu@yale.edu}{https://orcid.org/0000-0001-7291-3106}{}
\author{Fan Zhang}{Yale University, USA}{f.zhang@yale.edu}{https://orcid.org/0000-0002-8525-4514}{}
\author{Michael K.\ Reiter}{Duke University, USA}{michael.reiter@duke.edu}{https://orcid.org/0000-0001-7007-8274}{}
\authorrunning{L. Zhou, Z. Liu, F. Zhang and M.\ K.\ Reiter}

\Copyright{Lulu Zhou, Zeyu Liu, Fan Zhang and Michael K.\ Reiter}


\funding{This work was funded in part by NSF grant 2207214 and an Ethereum Academic Grant.}
\nolinenumbers

\EventEditors{Rainer B\"{o}hme and Lucianna Kiffer}
\EventNoEds{2}
\EventLongTitle{6th Conference on Advances in Financial Technologies (AFT 2024)}
\EventShortTitle{AFT 2024}
\EventAcronym{AFT}
\EventYear{2024}
\EventDate{September 23--25, 2024}
\EventLocation{Vienna, Austria}
\EventLogo{}
\SeriesVolume{316}
\ArticleNo{5}

\begin{document}

\maketitle

\theoremstyle{definition}
\newcommand{\rentingamt}{\mathsf{c_{cloud}}}
\newcommand{\insuranceamt}{\mathsf{c_{in}}}
\newcommand{\spcap}{\mathsf{c_{sp}}}
\newcommand{\amtcap}{\mathsf{v}}
\newcommand{\totalamtcap}{\mathsf{C_m}}
\newcommand{\compensationvalue}{\mathsf{v_c}}
\newcommand{\rewardvalue}{\mathsf{v_r}}
\newcommand{\disputepenalty}{\mathsf{p_d}}
\newcommand{\bountypenalty}{\mathsf{p_b}}
\newcommand{\keycap}{N}
\newcommand{\NN}{\mathbb{N}}
\newcommand{\txdispute}{\mathsf{tx\_disp}}
\newcommand{\zeyucamera}[1]{\dtcolornote[zeyu]{blue}{#1}}
\newcommand{\lulucamera}[1]{\dtcolornote[lulu]{orange}{#1}}
\newcommand{\zeyu}[1]{}
\newcommand{\lulu}[1]{}

\newcommand{\luluadd}[1]{{
		{\color{orange}
			#1
		}\normalcolor}}

\newcommand{\fanz}[1]{}
\newcommand{\mike}[1]{}
\newcommand{\publicdata}{\mathsf{p}}
\newcommand{\secretinputdata}{\mathsf{s}}
\newcommand{\rqtoken}{\mathsf{req\_token}}
\newcommand{\stateoauth}{\mathsf{state}}
\newcommand{\disputetime}{t'}
\newcommand{\ID}{\mathsf{ID}}
\newcommand{\id}{\mathsf{id}}
\newcommand{\rqtokens}{\mathsf{req\_tokens}}
\newcommand{\idlist}{\mathsf{id\_list}}
\newcommand{\provider}{\mathsf{pvd}}
\newcommand{\providerlist}{\mathsf{pvd\_list}}
\newcommand{\token}{\mathsf{token}}
\newcommand{\tokens}{\mathsf{tokens}}
\newcommand{\vrftokens}{\mathsf{vrf\_tokens}}
\newcommand{\txreq}{\mathsf{tx\_req}}
\newcommand{\replytxreq}{\mathsf{reply\_tx\_req}}
\newcommand{\idmap}{\mathsf{id\_mapping}}
\newcommand{\idprovider}{\mathsf{id\_pvd}}
\newcommand{\init}{\mathsf{init}}
\newcommand{\ZZ}{\mathbb{Z}}
\newcommand{\txinfo}{\mathsf{tx\_info}}
\newcommand{\source}{\mathsf{s\_addr}}
\newcommand{\dest}{\mathsf{d\_addr}}
\newcommand{\txvalue}{v}
\newcommand{\tkdb}{\mathsf{tx\_tokens\_db}}
\newcommand{\disputetkdb}{\mathsf{dp\_tokens\_db}}
\newcommand{\tx}{\mathsf{tx}}
\newcommand{\createtx}{\mathsf{creat\_tx}}
\newcommand{\signedtx}{\mathsf{tx_\sigma}}
\newcommand{\compensation}{\mathsf{c\_addr}}
\newcommand{\disputestring}{\mathsf{s_d}}
\newcommand{\keygenthreshold}{\mathsf{ThKeyGen}}
\newcommand{\signthreshold}{\mathsf{ThSign}}
\newcommand{\skshare}{\mathsf{ss}}
\newcommand{\ratecap}{\mathsf{c}}
\newcommand{\pkenclave}{\pk_{\mathsf{TEE}}} %
\newcommand{\pkattestation}{\pk_{\mathsf{Att}}} %
\newcommand{\query}{\mathsf{q}}
\newcommand{\querytee}{\mathsf{q_{t, \sigma}}}
\newcommand{\skenclave}{\sk_{\mathsf{TEE}}} 
\newcommand{\skattestation}{\sk_{\mathsf{Att}}} %
\newcommand{\encdisptokens}{\ct_{\mathsf{disp\_tk}}}
\newcommand{\dtokens}{\tokens_{\mathsf{d}}}
\newcommand{\result}{\ensuremath{r}}
\newcommand{\timelimit}{\ensuremath{t}}
\newcommand{\signature}{\ensuremath{\mathsf{sig}}}
\newcommand{\abstractmodel}{TEE-SC\space}
\newcommand{\abstractprotocol}{\text{avail}\space}
\newcommand{\utility}[1]{\ensuremath{U_{#1}}}
\newcommand{\valueofutility}[1]{\ensuremath{v_{#1}}}
\newcommand{\function}[2]{\ensuremath{f_{#1}(#2)}}
\newcommand{\bountynumshares}{\mathsf{k}}
\newcommand{\bountynumsharesvalid}{\mathsf{k'}}
\newcommand{\depositvalue}{\mathsf{v_d}}
\newcommand{\transactionfee}{\mathsf{v_f}}
\newcommand{\txbounty}{\mathsf{tx\_bounty}}
\newcommand{\bountydest}{\mathsf{b\_addr}}
\newcommand{\opoverhead}{\mathsf{c_{op}}}
\newcommand{\actreq}{\mathsf{act\_req}}
\newcommand{\replyactreq}{\mathsf{reply\_act\_req}}
\newcommand{\insuranceclaim}{\mathsf{insur\_req}}
\newcommand{\replyinsuranceclaim}{\mathsf{reply\_insur\_claim}}
\newcommand{\replybountyclaim}{\mathsf{reply\_bounty\_claim}}
\newcommand{\abstractteereply}{\mathsf{R_{r, \sigma}}}
\newcommand{\insurcontractq}{\mathsf{on\_recieve\_q}}
\newcommand{\insurcontractr}{\mathsf{on\_recieve\_r}}
\newcommand{\timeoutfunction}{\mathsf{on\_timeout}}
\newcommand{\hash}{h}
\newcommand{\bountyhash}{\mathsf{h_b}}
\newcommand{\sender}{\mathsf{sender}}
\newcommand{\txisdisputed}{\mathsf{disputed}}
\newcommand{\txisrecorded}{\mathsf{recorded}}
\newcommand{\txtokenisvalid}{\mathsf{tx\_vld}}
\newcommand{\penaltyvalue}{\mathsf{v_{p}}}
\newcommand{\numberofimplementations}{\ensuremath{N}}
\newcommand{\probabilityofbrokenimplem}{\ensuremath{p_i}}
\newcommand{\replyquerycost}{\mathsf{c_{exe\_req}}}
\newcommand{\replyqueryreward}{\mathsf{r_{exe\_req}}}
\newcommand{\reputationcost}{\mathsf{c_{rep}}}
\newcommand{\servicefee}[1]{v_{s_{#1}}}
\newcommand{\utilitygainfromservice}{{g_s}}
\newcommand{\lossofsecretmisuse}{{v_l}}
\newcommand{\keygen}{\mathsf{KeyGen}}
\newcommand{\sk}{\mathsf{sk}}
\newcommand{\pk}{\mathsf{pk}}
\newcommand{\ct}{\mathsf{ct}}
\newcommand{\Enc}{\mathsf{Enc}}
\newcommand{\Dec}{\mathsf{Dec}}
\newcommand{\vk}{\mathsf{vk}}
\newcommand{\sgk}{\mathsf{sgk}}
\newcommand{\sign}{\mathsf{Sign}}
\newcommand{\verify}{\mathsf{Verify}}
\newcommand{\tsign}{\mathsf{ThresholSign}}
\newcommand{\vrfsig}{\verify}
\newcommand{\Setup}{\mathsf{Setup}}
\newcommand{\ProofGen}{\mathsf{ProofGen}}
\newcommand{\pp}{\mathsf{pp}}
\newcommand{\sset}{\mathsf{sset}}
\newcommand{\adv}{\mathcal{A}}
\newcommand{\ext}{\mathcal{E}}
\newcommand{\negl}{\mathsf{negl}}
\newcommand{\pok}{\mathsf{PoK}}
\newcommand{\SC}{\mathsf{SC}}
\newcommand{\RR}{\mathbb{R}}
\newcommand{\updateshares}{\mathsf{UpdateS}}
\newcommand{\keys}{\mathsf{keys}}
\newcommand{\totalnumberofsecrets}{\mathsf{N}}
\newcommand{\sshare}{\mathsf{SecretShare}}
\newcommand{\recover}{\mathsf{Recover}}
\newcommand{\thresholdmat}{\mathsf{m}}
\newcommand{\operationcost}{c_o}
\newcommand{\maxreward}{\mathsf{C_w}}
\newcommand{\numusers}{\mathsf{U}}
\newcommand{\totalwallets}{\mathsf{W_{tot}}}
\newcommand{\maxwalletsperuser}{\mathsf{W}}
\newcommand{\claimedvalue}{\mathsf{v_{cl}}}
\newcommand{\snc}{\textsf{CrudiTEE}\xspace}
\newcommand{\scwallet}{\mathsf{SC}_{\text{wallet}}}
\newcommand{\scbounty}{\mathsf{SC}_{\text{bounty}}}
\newcommand{\scstick}{\mathsf{SC}_{\text{stick}}}
\newcommand{\smartcontractTEESC}{\mathsf{SC_{\abstractprotocol}}}
\newcommand{\smartcontractwallet}{\mathsf{SC_{\text{wallet}}}}
\newcommand{\smartcontractdata}{\mathsf{SC_{\text{data}}}}
\newcommand{\smartcontractinsurance}{\mathsf{SC_{\text{ins}}}}
\newcommand{\totaltimesteps}{T}
\newcommand{\psuccess}{p_s}
\newcommand{\costPerStep}{c_a}
\newcommand{\bountycap}{\alpha_\mathsf{cap}}
\newcommand{\costofdefender}{c_d}
\newcommand{\probsellingkey}{p_e}
\newcommand{\holdingtime}{t_h}
\newcommand{\fscore}{f}
\newcommand{\bonusoftime}{b_e}
\newcommand{\bonusofturnin}{b_t}

\newcommand{\thenewparagraph}[1]{\parhead{#1}}

\algdef{SE}[SUBALG]{Indent}{EndIndent}{}{\algorithmicend\ }%
\algtext*{Indent}
\algtext*{EndIndent}

\newcommand{\walletname}{\textsf{CrudiTEE}\xspace}

\begin{abstract}
Cryptocurrency introduces usability challenges by requiring users to manage signing keys.
Popular signing key management services (e.g., custodial wallets), however, either introduce a trusted party or burden users with managing signing key shares, posing the same usability challenges.
TEEs (Trusted Execution Environments) are a promising technology to avoid both, but practical implementations of TEEs suffer from various side-channel attacks that have proven hard to eliminate. 

This paper explores a new approach to side-channel mitigation through {\em economic incentives} for TEE-based cryptocurrency wallet solutions. 
By taking the cost and profit of side-channel attacks into consideration, we designed a Stick-and-Carrot-based cryptocurrency wallet, \snc\footnote{Crudite is a salad with carrots and (other) vegetable sticks.}, that leverages penalties (the stick) and rewards (the carrot) to disincentivize attackers from exfiltrating signing keys in the first place.
We model the attacker's behavior using a Markov Decision Process (MDP) to evaluate the effectiveness of the bounty and enable the service provider to adjust the parameters of the bounty's reward function accordingly.
\end{abstract}

\section{Introduction}
\label{sec:intro}

As cryptocurrencies~\cite{bitcoin,wood2014ethereum} gain popularity, more and more people use cryptographic signatures as a way to authorize transactions.
Unfortunately, signing key management has long been a notoriously hard problem. 
With inexperienced users often struggling with lost or leaked keys, a natural tendency is to outsource the task to specialized service providers.
For example, 11\% of the entire cryptocurrency marketization is stored in custody by a single service provider (Coinbase~\cite{coinbase11percent}).
This is undesirable security-wise, as the secrecy of keys (thus the safety of the funds) relies on the trustworthiness of a centralized party.

To provide stronger security guarantees (and to reduce liability), 
a cryptocurrency wallet service provider can generate users' signing keys in Trusted Execution Environments (TEEs, such as Intel SGX~\cite{anatiinnovative2013,mckeeninnovative2013}, 
AMD SEV~\cite{amdsev}, Nvidia H100~\cite{H100TensorCoreGPUNVIDIA}) and serve signing requests in TEE without ever seeing the signing keys in plaintext.
However, the naive adoption of TEEs does not provide a meaningful secrecy guarantee {\em to users}, because the service provider may be able to exfiltrate signing keys through side-channel attacks~\cite{nilsson2020survey}.
While side-channel mitigation has been extensively studied in the literature (e.g., see  \cite{szefer2019survey} for a survey), side channels are notoriously hard to eliminate, due to the complexity of modern processor design (e.g., TEEs often share physical resources with untrusted processes, such as caches). 

Our work is motivated by the observation that the operator of TEEs is the primary actor capable of mounting side-channel attacks, since most attacks \cite{wang2017leaky, nilsson2020survey, li2021cipherleaks,teeroot1,teeroot2,teeroot3} require root access to the host.
For wallet key management services, the TEE operator is the service provider. 
This observation gives us additional leverage to prevent side-channel attacks, as the service provider can be held responsible (using techniques to be presented later) if a wallet key is leaked or accessed without user authorization.
Using a proper penalty mechanism, we can eliminate the service provider's gains from a successful side-channel attack, thus removing the incentive to attack in the first place. 

With the TEE operator striving to avoid key leakage, the possibility of side-channel attacks by non-local, unprivileged attackers is significantly reduced (e.g., the service provider is motivated to employ heightened security measures). To further discourage such attacks, our idea is to reward the attackers for partial success. For example, if a signing key is distributed cross $\keycap$ TEEs using secret sharing, we give the attacker a substantial reward if he successfully exfiltrated {\em any} share. With a proper reward function, this early reward can serve as a strong incentive for the attacker to {\em stop early}, giving the system administrator time to react to partial compromise before a full key is exfiltrated.

\subsection{\snc: The Cryptocurrency Wallet with Stick and Carrot}
\label{sec:intro:snc}

Based on the above two principles, we propose \snc, 
a TEE-based cryptocurrency wallet that can defend against TEE side channels by privileged and unprivileged attackers, using penalties (stick) and rewards (carrot), respectively.
Furthermore, \snc strives to achieve user-friendliness (i.e., users do not need to store keys locally). 
\snc first requires that the signing keys be generated inside TEE and never exported in plaintext. Assuming correct implementation, this implies that signing key leakage is impossible except for through side-channel attacks. 

We classify potential actors capable of mounting side channel attacks into {\em insider attackers} and {\em outsider attackers}.
The insiders are privileged attackers, such as service providers, who have full control over the TEE including physical access.
Insiders have powerful attacking capabilities required by most side-channel attacks (such as root privilege) like the ones needed in ~\cite{kocher1999power,gandolfi2001electromagnetic, quisquater2001electromagnetic}.
In contrast, the outsiders are all the attackers who can exfiltrate the secrets in the TEEs only through less-privileged means like remote time-based attacks \cite{kocher1996timing, brumley2005practical, aciicmez2007remote}. We refer readers to \cref{sec:related:side-channels} for more examples.  
As introduced above, \snc consists of the stick (penalties), to discourage insider attackers, and the carrot (rewards), to encourage outsider attackers to stop early.

Note that to perform such punishment or distribute the bounty,
we need an automated but also trustworthy and publicly accessible mechanism. Smart contracts~\cite{wood2014ethereum}(autonomous programs executed on blockchains) are the perfect tool for this purpose.
Thus, below, when discussing the stick and the carrot, we use the smart contract as an important building block.

\subsubsection{The stick}
\label{sec:intro:the stick}
Due to the power of the service provider, 
preventing it from mounting side channels via a technical way seems infeasible.
Instead, \snc requires the service provider to put down {\em collateral}, which will be confiscated if signing keys safeguarded by the TEEs are used for unauthorized signatures or if legit service requests from users are denied. 

To realize the stick of \snc, the key is to enable a user to generate publicly verifiable proof if her TEE-generated keys are illegally accessed. 
First, as mentioned, raw keys stay in the TEE and are never exported outside.  
Second, each key corresponds to a wallet owner and can only be used by the owner through well-defined APIs (e.g., an API could allow the owner to sign messages with the key using a carefully implemented signature algorithm).  
Third, to access a key, a signed authorization from its owner must be present and checked by TEEs, thus making the authorization process \textit{accountable} (i.e., if the user disputes a signature, the service provider can present proof that the signature was authorized by the user).
Users can verify TEE attestations to ensure the prerequisites are met before signing up for the service.

In order not to burden the user with signing key management while making the authorization process accountable,
we use the OAuth protocol (\cref{primitive:oauth}).
The token signed by the OAuth provider is used as proof of authorization.

The service provider sets up a smart contract to implement the insurance 
(denoted
$\smartcontractinsurance$%
) with the following logic and makes an initial deposit. 
If a user discovers any unauthorized signature, she can submit a request to $\smartcontractinsurance$. The service provider must prove that the user had authorized such key use within a specific period. Failing to provide such proof results in the insurance smart contract automatically compensating the user.

\subsubsection{The carrot}

Without the help of any insiders, outside attacks become unlikely, but still not impossible.
To limit potential exposure to external attacks, we employ the threshold signing protocol such as~\cite{herzberg1995proactive}, where the signing key is stored as key shares across multiple independent TEEs (e.g., hosted in different clouds) and refresh secret shares periodically. This way, even if an outside attacker can exfiltrate a few shares, he needs all shares to exfiltrate the entire key.
However, the security of such proactive secret sharing method as a defense is ``black or white'' --- unless the attacker can break a sufficient number of TEEs and cause a catastrophic breach, partial breaches cannot be detected and therefore cannot inform the service provider to take proper action to prevent those catastrophic breaches.
By exploiting economic incentives, we can elicit such information from the attacker. Specifically, \snc enhances a proactive secret-sharing scheme with an alerting mechanism so that when partial breaches happen (e.g., TEEs deployed in one cloud are vulnerable, but not others),
the attacker is encouraged to alert the service provider in exchange for a {\em bounty}. This allows the service provider to take proper action before full breaches happen. 

Designing a bounty reward function that induces the desired behavior of the attackers is the main technical challenge.
Specifically, we aim to formulate a reward function that motivates attackers to promptly alert the service provider without generating any illegal signature or selling the acquired signing key shares, while minimizing the defender's cost (i.e., the service provider's cost).
We employ a 2-step methodology in the reward function design: 
we start with the attacker with a fixed known cost first and then deal with the one whose attacking process is non-deterministic and whose cost cannot be accurately estimated.

\parhead{Step 1:}
We provide the following toy example to illustrate the challenge in reward function design under a deterministic setting.
We start with a key (worth \$3 in total) stored as three secret shares, each of which is worth \$1 (assuming a share can be sold on the market for \$1).
To steal one share, the attacker's cost is \$0.8.
Furthermore, assume that \$0.01 is the smallest unit of money for simplicity.
Without a bounty, the attacker will keep attacking until he gets 3 shares and sells them on the market for \$3, making a profit of \$0.6.

To protect against such an attacker, there are two naive but natural solutions.
The first solution is to simply have the reward function be a constant function of \$3.01 (i.e., the attacker obtains \$3.01 for any amount of shares he steals).
In this case, an attacker always submits the share as soon as he obtains the first share,
but then the defender costs more than the key value itself.
The second solution is setting the function to be \$1.01 per share (i.e., a function linear in the number of shares).
However, in this case, an attacker would instead try to obtain all three shares and claim a total reward of \$3.03, which costs even more.

The optimal solution is to set the reward function to be a constant function of \$1.41 (i.e., the attacker is awarded \$1.41 for finding any amount of shares): 
the attacker will stop attacking and turn in the key shares whenever he obtains 1 key share, making a profit of \$0.61.
This reward function not only encourages the attacker to submit as soon as getting one share but also minimizes the defender's cost. 
Note that it is indeed the least the service provider can pay,
as if the reward is less than \$1.41, the attacker will sell the key for a higher profit instead (assuming w.l.o.g. that the attacker sells the secret when the profit from the bounty is tied with selling the key).

\parhead{Step 2:}
The reward function in the toy example, however, is based on a simplified assumption of deterministic attack costs and requires the defender to accurately know the attacker's cost. Our design instead aims to address real-world situations where the attacker's attack process is non-deterministic, and the cost of attacking cannot be accurately known in advance.

To design the reward function in this setting,
we first turn the desired properties of the reward function into numerical metrics. 
Then we capture the non-deterministic attacking process as an ``optimal stopping'' game and use Markov Decision Process (MDP) to analyze the attacker's optimal strategy.
We propose a reward function for non-deterministic attackers and optimize it using the metrics as an objective function, based on the defender's budget and estimation of the attacker's cost and success rate.
We further show that the reward function not only has good performance for the attacker with an accurately estimated cost but also for attackers with different costs.
We provide the defender with the performance of the optimized reward function for attackers with a wide range of costs and success rates. 
The defender can use such a strategy to assess how the reward function she obtains performs for a range of attackers.
If she is not satisfied with the result,
she could raise their budget and generate another function.

To realize the bounty, the service provider creates a smart contract $\scbounty$ that accepts proofs of knowledge (PoK) of TEE-managed key shares and remits rewards accordingly.
Valid PoK submissions to $\scbounty$ raise a flag, pausing operations until the keys are rotated and the flag is reset. 
To ensure that the attacker did not use the breached key for unauthorized signings, users are requested to check for unauthorized signatures during the shutdown period. 
If any are found, the attacker's reward is forfeited. 

\subsection*{Contribution}

We summarize our contributions as follows: 
\begin{enumerate} 
    \item We introduce a new approach to building a cryptocurrency wallet: \snc that
    leverages \textit{economic incentives} to defend against side-channel attacks from insiders and outsiders.
 
    \item \snc involves a novel automatic insurance system (\cref{sec: stick}), allowing users to receive compensation if their wallet signing key is used for signing transactions without their authorization. 
    
    \item We develop a reward function for the bounty in \snc (\cref{sec:bounty}) that encourages attackers to submit key shares to the bounty immediately while minimizing the defender's cost. We use the Markov Decision Process (MDP) to model the non-deterministic nature of side-channel attacks and optimize the reward function against numerical metrics. We evaluate and show the optimized reward function is effective not only for attackers with precisely estimated costs but also for attackers with variable costs.
    The service provider may adjust her budget to cover a wider range of attackers the reward function can effectively defend against based on the evaluation.
    
\end{enumerate}

\section{Related Work}
\label{sec:related}

\subsection{Cyber Bounty}
Setting up bug bounties is a popular way to defend against hackers \cite{malladi2019bug}.
However, a fair exchange of bugs and money is difficult without trust.
Breidenbach et al.~\cite{breidenbach2018enter} proposed that
smart contracts to be deployed to guarantee that the attacker gets paid once a valid bug is submitted.
Their game-theoretic analysis showed that the attacker is incentivized to submit the bug as soon as possible because of competition from other honest hackers.
However, this is not always the case for side-channel attacks:
a malicious attacker may be the only one to discover a zero-day\footnote{A zero-day is a venerability in software or hardware that is unknown to its vendor.} side channel.
That is why we take the submission time into consideration in our reward function, i.e., to incentivize attackers to submit the leaked signing key (share) immediately upon acquiring it.

\subsection{Side Channels}
\label{sec:related:side-channels}
Side-channel attacks against cryptographic systems usually take one of three forms.  \textit{Time-driven}
side-channel attacks expose key information by monitoring \textit{total} execution times of cryptographic operations with a fixed key, which can reflect interactions among the value of the key, the structure of
the cryptographic implementation, and system-level effects such as cache evictions (e.g.,~\cite{kocher1996timing, brumley2005practical, aciicmez2007remote, weiss2012cache}).  \textit{Trace-driven} side-channel attacks observe a time-series signal reflecting a device's cryptographic operation \textit{throughout its execution}, e.g., by monitoring the device's power draw during the operation (e.g.,~\cite{kocher1999power}) or its electromagnetic emanations (e.g.,~\cite{gandolfi2001electromagnetic, quisquater2001electromagnetic}).  Finally, in an \textit{access-driven} side-channel attack, the attacker executes a program on the same computer where the cryptographic operation is taking place, using this vantage point to monitor the operation's use of microarchitectural components on the platform (e.g.,~\cite{osvik2006prime, gullasch2011games, gruss2016flush}).  Time-driven and trace-driven attacks are largely agnostic to the encapsulation of the cryptographic operation within a TEE.  In contrast, much effort has been expended to adapt access-driven attacks to attack a cryptographic operation executed within TEE from outside, with considerable success (e.g.,~\cite{wang2017leaky, nilsson2020survey, li2021cipherleaks}).

Using the terminology of \cref{sec:intro}, we consider \textit{outsiders} to be less privileged and thus limited to time-driven and some access-driven attacks, that can be performed remotely (i.e., without any physical access to the TEE).  
Any attacks available to an outsider, however, must incur costs to conduct over time, e.g., to achieve and maintain co-residency on the same physical computer as the victim computation~\cite{varadarajan2015placement} (possibly despite defenses to make this difficult, e.g.,~\cite{moon2015nomad}) and to perform attack computations.  
In contrast, \textit{insiders} are permitted to conduct \textit{any} time-driven, trace-driven, or access-driven attacks, and so are considerably more powerful.  In particular, we design \snc in anticipation of insiders capable of extracting keys from TEEs easily.  Outsiders, on the other hand, are assumed to require more time and costs to mount their attacks.

\subsection{TEE Side-channel Defense}
A recent concurrent and independent work, Sting \cite{babel2023sting}, proposes to use SC as a bug bounty, 
which is set up to encourage anyone who has access to a leaked secret to submit proof.
The proof of leakage is acquired in this way: 
first, a prover-owned TEE generates a secret, without disclosing it to the prover. 
Second, the secret is directly sent to the secret management service provider (without exposing the secret to the prover).
Finally, the prover acquires the secret using a side-channel attack, sends it back to the prover-owned TEE, and gets a proof of leakage from the TEE.
Sting focuses more on the proof generation rather than the bounty design,
however.
This is different from our bounty as we encourage attackers (\textit{without} physical access to the machine) to stop recovering the secret and submit a bounty claim without recovering the whole secret via economic incentives.

Numerous techniques other than bug bounty could be applied to side-channel defense, including ORAM \cite{costa2017pyramid}, code hardening \cite{brickell2006mitigating}, data location randomization \cite{brasser2019dr}.
However, defenses introduce performance overheads and usually defend against only specific types of attacks.
Another problem is that a service provider might not have enough incentive to apply these defensive technologies expeditiously.
Therefore, motivating the service providers to keep their TEEs safe from attack is crucial to the real-world use of TEEs.

\subsection{Existing Wallet Solutions}
\label{sec:existing wallet and CA solutions}
Some companies provide the service like a centralized bank for cryptocurrency \cite{coinbase}, holding users' funds in company-owned accounts.
Such centralized service deviates from the decentralized nature of cryptocurrency and increases risk to user funds.
On the other hand,
there are products to enable users to store their signing keys in a protected area of an offline device, named hardware wallet \cite{suratkar2020cryptocurrency}.
This approach raises costs and complicates transactions, and users usually have to trust the software provided by the hardware manufacturer for signing transactions.
A keyless wallet was constructed using witness encryption \cite{zindros2021hours}. 
To access the money, the user only needs to provide a short one-time password of 6 alphanumeric characters generated from an offline device.
Since Witness Encryption is currently impractical, however, the scheme is largely theoretical. 

\section{Background and Preliminaries}
\label{sec:prelim}

\subsection{Trusted Execution Environments}
\label{subsetction:tee}

TEEs (Trusted Execution Environments) are secure and isolated execution environments that provide confidentiality and integrity guarantees and the ability for a party to remotely verify the status of a TEE through remote attestation. Prominent examples of TEEs include Intel SGX~\cite{anatiinnovative2013,mckeeninnovative2013}, 
AMD SEV~\cite{amdsev}, and Nvidia H100~\cite{H100TensorCoreGPUNVIDIA}.
A major practical limitation of TEEs is side channel attacks (\cref{sec:related:side-channels}) that could break the confidentiality guarantee.

\subsection{Smart Contracts}
\label{subsection:blockchain_and_smartcontract}

To create elaborate economic incentive structures, \snc uses smart contracts, autonomous programs running on top of blockchains, to remit payments under specific events.
We follow the standard assumption that smart contracts are correct (i.e., the security assumptions required by the blockchain protocol are met) and available (i.e., all parties in our protocols can access the smart contract and request submitted to the smart contract is executed within a time limit).

\subsection{OAuth}
\label{primitive:oauth}

\snc uses the OpenID Connect feature in OAuth (Open Authorization) 2.0 \cite{GoogleOIDC,oauth2} to enable users to make signing requests without possessing a signing key. 
OpenID is an authentication protocol that allows users to use an existing account from an OpenID provider (denoted as ``OAuth provider''), such as Google, to authenticate themselves on other applications. Furthermore, during authentication, a user can embed a customized message in the `nonce' field of the signed ID token~\cite{GoogleOIDC} (looking ahead, this allows the user to put a description of her request in this field).

\subsection{Cryptographic Primitives}

We provide a brief description of the threshold signing scheme. Formal definitions of other cryptographic primitives are presented in~\cref{sec:formalprimitive}. 

Threshold signature allows $\keycap > 1$ parties to share a secret signing key, such that each party obtains a share of the signing key.
Only when $m$ parties owning a sharing, $1 \leq m < \keycap$, together can sign a message. 
Knowledge of $< m$ shares leaks no information about the secret signing key.
Furthermore,
when the secret shares are updated to $\keycap$ new shares,
even $m_1 < m$ old shares and $m_2 < m$ new shares where $m_1 + m_2 \geq m$
together leak no information about the secret.
We use it to allow multiple TEEs to share the signing key,
such that only if $\geq m$ shares are leaked, the secret is leaked.

\subsection{Markov Decision Process}
A Markov decision process (MDP) is a mathematical model that captures decision-making under uncertain situations.
A Markov state is a state $S_t$ at time $t > 0$ satisfying $\Pr[S_t | S_{t-1}] = \Pr[S_t | S_{t - 1}, \dots, S_1]$ (i.e., the previous state captures the entire history states).
The MDP consists of a sequence of Markov states and an associated state transition matrix. This matrix represents the probabilities of transitioning from one state to another based on the player's actions. 
The player's optimal strategy in MDP can be computed using tools like \cite{chades2014mdptoolbox}.

\section{Threat Model and Roadmap}
\label{sec: overview}

\subsection{Threat Model}
The purpose of the techniques in \snc is to mitigate the side-channel attacks that break the privacy of the TEEs but not the integrity.
We assume TEE integrity (i.e. the data and code in the TEE cannot be modified by any attacker) to hold and remote attestation to be secure, following a common assumption (c.f.,~\cite{sgp,cheng2019ekiden}), as the attestation key is only used through a limited interface, unlike application-generated secrets.
The side-channel attacks that are strong enough to compromise the attestation key \cite{sgaxe} are out of scope for this work, 
as such incidents have historically been rare.

We assume that the integrity and liveness of smart contracts are enforced by the blockchain.
Furthermore, we assume the OAuth providers are trusted,
but note that any user can choose her own set of OAuth providers to trust (i.e., the user can choose a subset of a predefined set of OAuth providers).
Finally, we assume that both the service provider and the outsider attacker are rational entities aiming to maximize their profits. 
We do not consider non-financial incentives, and the agent who attack the system as a mere malicious intruder is out of our scope.

\subsection{Wallet Design Overview}
\label{subsection: overview of the framework}
In our wallet service, each client registered with the wallet service provider has a wallet whose signing key is stored in the service provider's TEE.
Our goal is to defend side-channels against such signing keys.

We categorize side-channel attacks into two types: insider attacks, which require physical access and/or root privileges, and outsider attacks which can be executed remotely without such privileges (\cref{sec:related:side-channels}).
In our wallet design, the service provider, who controls the TEEs, is classified as an insider, whereas all other attackers, including users, are categorized as outsiders. 
We defend the insiders using the insurance (the stick) and the outsiders using the bounty (the carrot).

The side-channel mitigation in \snc thus consists of three main components: 
\begin{enumerate} 
\item The accountable signing key management service (\cref{subsection: Authentication and Signing Key Management}) enables the users to register for the service and authorize the service provider to sign a transaction when needed.
\item The insurance (\cref{sec:stick}) ensures the service provider provides the desired service, and otherwise is punished. 
\item The bounty (\cref{sec:bounty}) aims to incentivize the outsider attacker to submit the key shares acquired through the remote side channel to the bounty (smart contract) rather than using them to make unauthorized signatures or selling them. 
\end{enumerate}

Both the insurance and the bounty are initiated using smart contracts ($\smartcontractinsurance$ and $\scbounty$). 
In addition, to make sure that the service provider answer all the service requests (instead of ignoring those requests),
the smart contract $\smartcontractTEESC$ is also deployed.
During setup, the service provider needs to build the TEE program and publish the attestation.
Then, the service provider deploys the aforementioned smart contracts 
on the blockchain.

To use the service, 
the user first chooses the OAuth provider(s) she trusts and creates a new account with her OAuth token (signed by that OAuth provider(s)). 
The service provider will execute the threshold key-generation protocol among the TEEs, register the OAuth account and key mapping, and then provide the public key to the user.
It is essential that the signing key is generated within the TEEs and remains within the TEEs (i.e. cannot be exported in plaintext format). 
This is because if the users learn the key, it becomes ambiguous whether the responsibility for any unauthorized signature lies with the users or the service provider.
After the generation of the signing key,
a smart contract wallet $\scwallet$ will be deployed for the user.
$\scbounty$ will also be updated so that the new key is also protected by the bounty.
The proof-of-publication\footnote{Proof of publication is a way for the TEE to verify that a state change is updated on the blockchain.} scheme is employed to ensure that the smart contract update is done properly.

The service provider replies to the user's transaction signing requests with authentication via OAuth providers (\cref{fig:overview_framework1}).
The signing is conducted using the threshold signature scheme,
with the signing key secret-shared among several TEEs.
When the service provider is not responding to a signing request, the user can send the request through $\smartcontractTEESC$ and force the service provider to respond.
If the user realizes that an unauthorized signature exists, 
she can submit a claim to $\smartcontractinsurance$ and get compensated
(\cref{fig:overview_framework2}).

Finally, 
if an outsider attacker steals the signing key (shares) from a remote side channel, 
he can submit it to $\scbounty$ and get rewarded based on the submission time and number of shares he submits
(\cref{fig:overview_framework2}).
Any valid $\scbounty$ or $\smartcontractinsurance$ submission will trigger a flag to signify that some of the TEEs have been breached. 
\snc requires that all wallet transactions cease until the service provider rotates all the signing keys and clears the flag.
If the full key is leaked, the TEE will generate a fresh key pair, update the OAuth account and wallet key mapping, and transfer the money in the smart contract wallet to the new wallet while the red flag is on. Transactions during the red flag period can only be triggered by a message signed by the TEE attestation key.
The reward for the attacker will be held for a specified period, during which the user of the affected keys will be asked to check whether there exists any unauthorized transactions and the reward will not be given to the attacker if such transactions are found.

For more details about $\scwallet$, $\smartcontractinsurance$ and $\scbounty$, we refer the readers to \cref{appendix: details of insurance and wallet} and \cref{appendix: bounty}.

\subsection{Reward Function Design Roadmap}
The attacker’s reward is determined by a reward function designed to incentivize them to claim the bounty immediately upon obtaining a single key share from the TEEs, while minimizing the defender's cost (\cref{subsection: reward function design}). 
Since the reward function design is particularly challenging among other components of the wallet,
we discuss our roadmap here.
We employ a 2-step methodology here:
first, we deal with attackers with known deterministic costs (a simplified case),
then we employ the ideas from this simplified case together with other more advanced mechanisms to develop the reward function for the attacker with non-deterministic and unknown costs.

In more detail, we begin with a case study assuming the attacker operates under a deterministic cost function known by the defender.
However, in the real world, the side-channel attacking process is non-deterministic, and the cost of the attack is hard to estimate accurately.
Building on insights gained from the case study, we propose a reward function for attackers with non-deterministic behavior. 
We model the non-deterministic attacking process as the ``optimal stopping'' game \cite{optimalstopping1,optimalstopping2,optimalstopping3} and employ Markov Decision Processes to calculate the best strategies for the attackers.
By translating the desired properties of this reward function into quantitative metrics used as the objective function,
we optimize the parameters in the reward function (based on the defender's budget and her estimation of the capability of the attacker).
Finally, we evaluate the effectiveness of our proposed reward function when the attacker's ability (parameterized by his cost and success rate) is different from the estimations. 
Based on the evaluation of the attacker, the defender can further raise her budget and recompute the function to get a more satisfying range of attackers the function can defend against.

\section{The Stick}
\label{sec: stick}
In this section, we first provide more details about the wallet workflow (\cref{subsection: Authentication and Signing Key Management}), which outlines the responsibilities of the service provider.
Then, we specify the ``stick'' part which holds the service provider responsible (\cref{sec:stick}).

\subsection{Authorization and Signing Transactions}
\label{subsection: Authentication and Signing Key Management}
We start by elaborating on how we make the authorization of the transactions accountable and describe how a user registers for an account and requests signed transactions.

\parhead{Accountable authorization}
\label{subsubsection: Accountable authentication}
As mentioned in the \cref{sec:intro:the stick}, an authorization process is \textit{accountable} if it leaves a signed evidence that can be used to prove the validity of the signing key usage later.
Meanwhile, it should not burden the user with additional key management.

Our solution leverages a feature in OAuth 2.0 called OpenID Connect (OIDC)~\cite{oauth2,GoogleOIDC}.
Specifically, OIDC-enabled OAuth providers issue signed identity tokens (called \texttt{ID\_token} \cite{GoogleOIDC}) that include a user identifier (such as email addresses) and a nonce set by users. 
Many mainstream OAuth providers enable the user application to specify the nonce in the ID token (e.g., Google \cite{GoogleOIDC}, Microsoft\cite{MicrosoftIDtoken}, etc.).

Every time the signing key is used,
we require the user to provide an ID token signed by the OAuth provider(s), which is uniquely linked to that specific signing request by including the request hash in the nonce field.
TEE verifies the token of the corresponding OAuth provider(s)' keys accordingly.
The public key of the OAuth providers is hardcoded in TEE and verified by the user through attestation.
This method not only provides a log-in process that most users are familiar with,
but also delegates authorization to a third party (or a set of third parties) that they trust, 
providing signed OAuth token(s) as proof of authorization.

\begin{figure}[t]
  \centering
  \subfloat[\centering Registration process.]{\includegraphics[page=1,width=0.5\columnwidth]{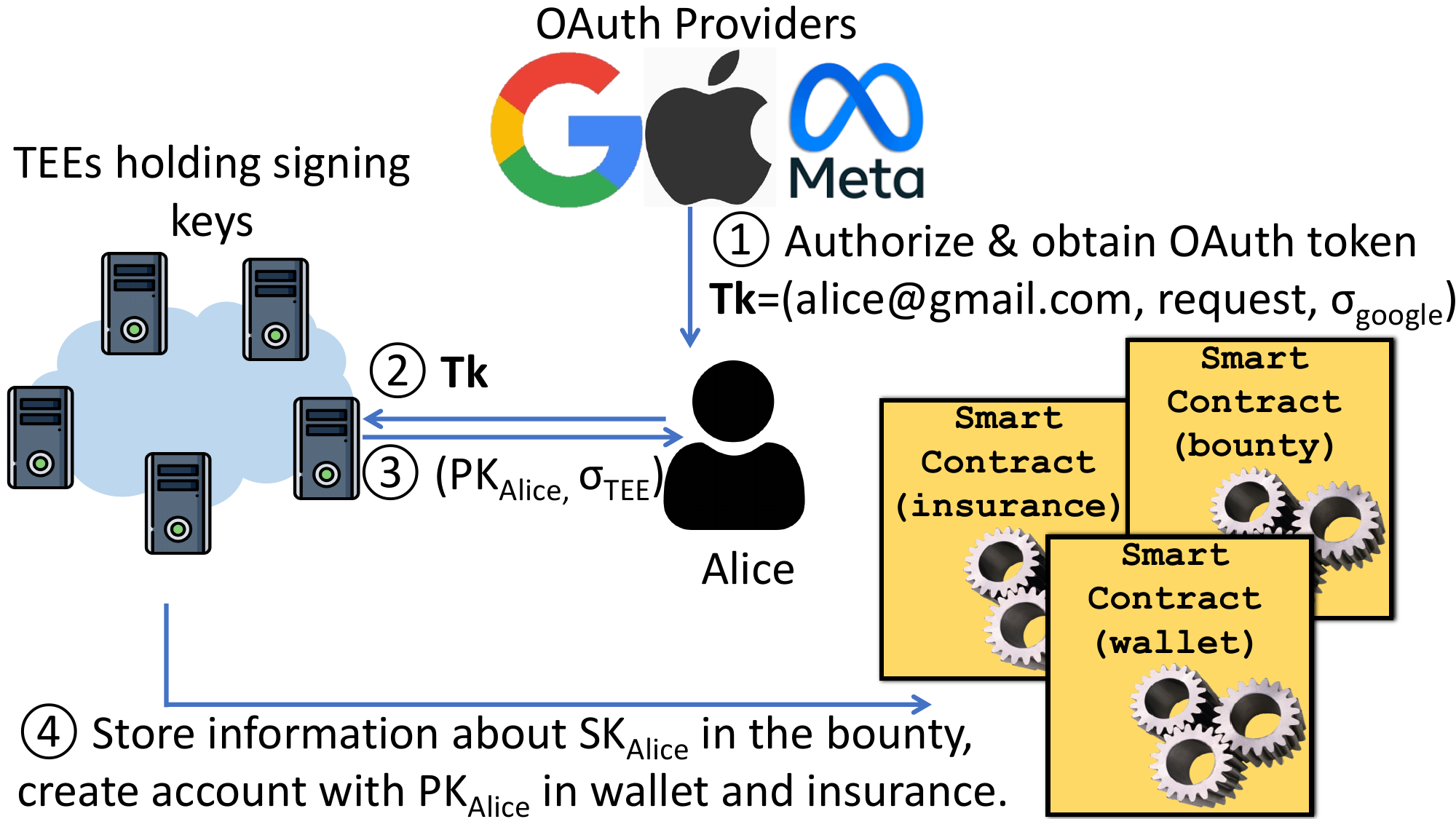}}
  \subfloat[\centering Transaction signing process.]{\includegraphics[page=2,width=0.5\columnwidth]{figures/digrams.pdf}}
\caption{Registration and Transaction Signing Workflow}
  \label{fig:overview_framework1}
\end{figure}

\parhead{Registration}
As shown in \cref{fig:overview_framework1} (a), when registering for a new account, the user runs a protocol to determine the future authentication process with the service provider.
Specifically, the user first chooses a set of OAuth provider(s) she trusts.
Next,
she puts the hash of the account registration request (e.g. the hash of ``\snc account registration'') in the `nonce' field of the ID token, authenticates it with the OAuth provider, and asks the OAuth provider to sign it.
Then, the user sends the account registration request to the service provider along with the token(s).
TEE verifies the token(s) and generates a fresh key pair for signing.
The TEE creates a TEE-signed receipt with the newly generated verification key (to verify the signed transactions for this user's wallet) and the OAuth ID(s) associated with it.
Lastly, a smart contract wallet is created for the user.

\parhead{Transaction signing request}
As shown in \cref{fig:overview_framework1} (b), when the user wants to sign a transaction, 
she generates a signing request.
Then, she acquires a signed token from the OAuth provider(s) with the hash of the transaction included in the token(s).
Once receiving the signing request and token(s), the service provider should input it into the TEEs. 
The TEE will check the validity of the request by verifying the token(s) and respond accordingly (we discuss how to enforce the TEEs to respond in \cref{subsubsection: Ensure Functionality of TEE}).
If the request is valid, the TEE will reply with the signature of the transaction, generated with the signing key associated with the user's OAuth ID(s).
If not, the TEE will reply with a message saying that the request is invalid, signed with its attestation key.
We require TEEs to store the (valid) tokens and requests in case of any future insurance claim (\cref{subsubsection: Insurance for unauthorized signing}).
The signed transaction will be submitted by the user to the wallet smart contract $\scwallet$.
The wallet smart contract will check the signature and execute the transaction. 
The pseudocode of $\scwallet$ is given in \cref{appendix:wallet smart contract}.

\parhead{Threshold signing}
\snc use a threshold signature scheme (e.g., ~\cite{gennaro2018fast}) for singing.
Specifically, 
the key-management service provider secret-shares each key into $\keycap$ secret shares
using a $(m,\keycap)$-threshold-signature scheme (where $m \leq \keycap$), stores them in independent TEEs, and rotates them every $T$ units of time.  
This approach not only serves to complicate the execution of side-channel attacks but also establishes the foundation for the bounty scheme described in \cref{sec:bounty}.

\subsection{The stick: hold service provider responsible}
\label{sec:stick}

Based on the accountable signing process described in the previous subsection, the ``stick'' aims to establish mechanisms to punish the service provider when it misbehaves.
The goal is that \textit{any} rational service provider would not choose to misbehave (e.g., steal the secret and produce an unauthorized signature).

\subsubsection{Ensure Availability of TEE}
\label{subsubsection: Ensure Functionality of TEE}

We start by discussing how to ensure that service providers process requests using TEE (with the expected inputs), guaranteeing TEE's availability \footnote{The idea of using incentives to make a service available is not new, though.
A similar method is used in blockchain Layer2 to prevent transaction censorship \cite{arbitrumsequencer}.
}. 
The service provider sets up $\smartcontractTEESC$ and makes the initial deposit.
If the service provider refuses to process a signing request directly submitted to the service provider,
the user submits the request to $\smartcontractTEESC$.
The service provider monitors the SC, processes any request from the SC, and forwards the request to the TEE. The TEE then generates a reply, which is either the requested signature or indicates that the request is invalid.
The reply, along with the user's request, must be signed by the TEE's attestation key.
After receiving the reply,
$\smartcontractTEESC$ checks whether the reply is signed by the TEE's attestation key and the request is included in the signed message.\footnote{Attestation key is hardcoded to the smart contract.}
If it is, $\smartcontractTEESC$ records the reply.
If the service provider does not submit a valid reply within a time limit, its deposit gets burnt (destroyed).
The workflow graph of $\smartcontractTEESC$ can be found in \cref{appendix:TEE Availability}.
\footnote{Note that one may consider a DoS-attack: initiating many small transactions using $\smartcontractTEESC$.
To avoid this, the service provider can setup a corresponding transaction fee to use $\smartcontractTEESC$ paid by the user.
If the user, however, needs to use such a service,
the user may consider the service provider as malicious, thus withdrawing all the money and stop using the service.
Thus, a rational service provider would avoid letting the user make transactions via $\smartcontractTEESC$.}

\subsubsection{Insurance for unauthorized transactions}
\label{subsubsection: Insurance for unauthorized signing}

\begin{figure}[t]
  \centering
  \subfloat[\centering Insurance Claim through $\smartcontractinsurance$.]{\includegraphics[page=4,width=0.5\columnwidth]{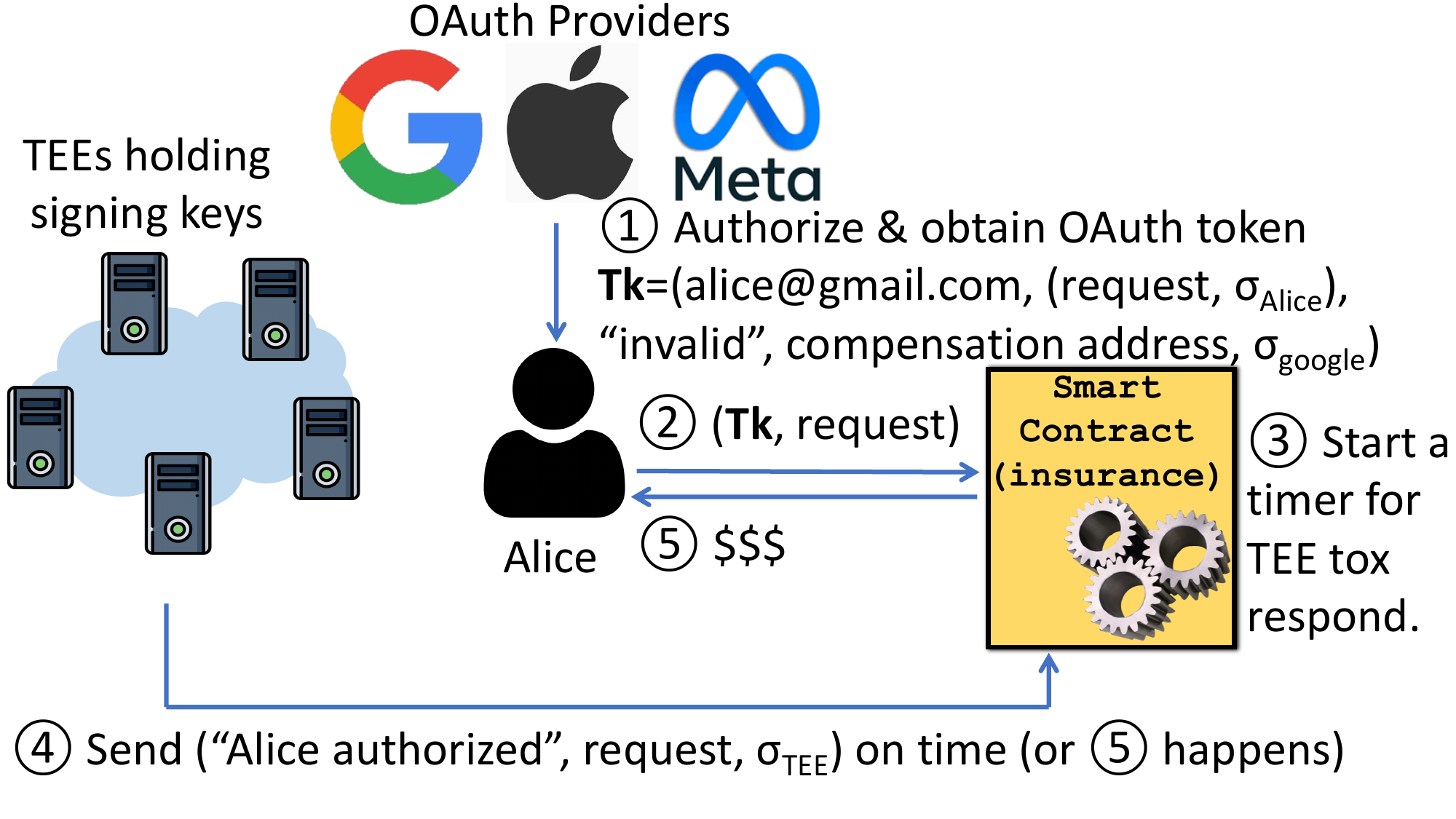}}
  \subfloat[\centering Bounty claim through the $\scbounty$.]{\includegraphics[page=5,width=0.5\columnwidth]{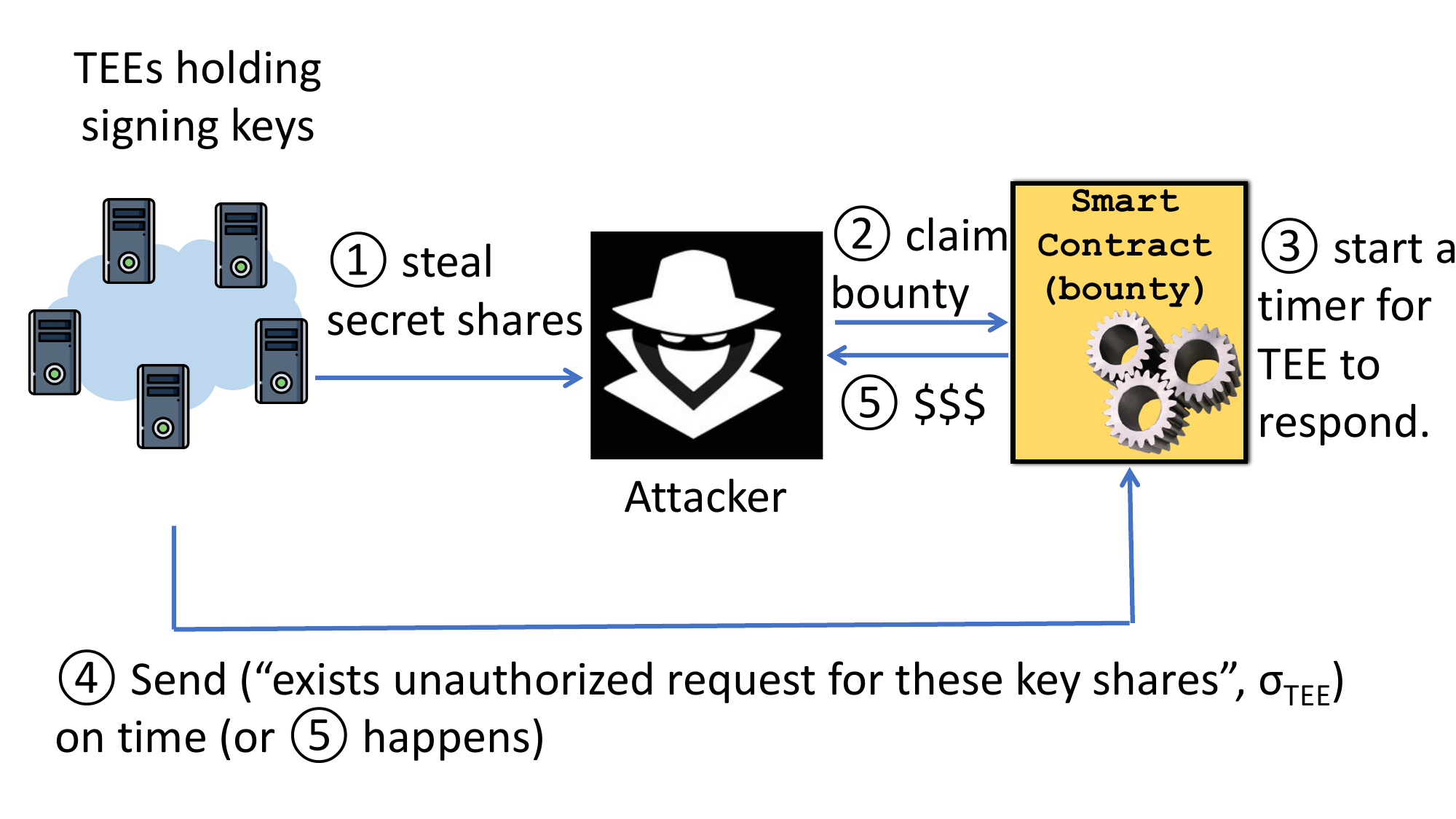}}
\caption{Insurance and bounty workflow}
  \label{fig:overview_framework2}
\end{figure}

In this part, we develop a mechanism that enables users to report unauthorized transactions.
As shown in \cref{fig:overview_framework2} (a), the user submits the signature to request a message, signed by the TEE's attestation key, stating that the signature is authorized by the user.
When the service provider is unable to provide such a message, the user is automatically compensated.
Since the user initiates the insurance claim, they are responsible for monitoring transactions and submitting complaints for unauthorized transactions, similar to most systems based on staking and slashing~\cite{li2023security}.

We instantiate the insurance using a smart contract ($\smartcontractinsurance$).
This smart contract specifies the necessary ground truth requirements, such as the attestation key of the TEEs, and the conditions under which users are eligible for compensation.
A predefined quantity of deposits is deposited in it, serving as potential compensation for the user.

An insurance claim is initiated by the submission of an unauthorized transaction to $\smartcontractinsurance$ together with the proof of ownership of the key.
The proof of ownership is a message stating the ownership of the key signed by the TEE, which could be requested using the user's OAuth token.
$\smartcontractinsurance$ checks whether the claim for the transaction has not yet been made before.
If yes, the claim will be rejected.
The service provider monitors $\smartcontractinsurance$ and sends the request to the TEE once it is published on the blockchain.
The TEE looks for the authentication token(s) associated with this request
(recall that the valid requests are stored).
If no valid token(s) in question are found, 
the TEE will sign a message stating that the signature was unauthorized with its attestation key.
Otherwise,
a message stating that the signature was authorized will be signed.
The service provider submits the reply to $\smartcontractinsurance$.
$\smartcontractinsurance$ checks whether the message signed by the TEE attestation key states that the signature was authorized.
If not, $\smartcontractinsurance$ compensates the user (for some predetermined value that depends on the application) and records this claim (e.g., on the chain) for future reference.
If the service provider fails to submit the requisite proof within the specified timeframe, 
the user automatically gets compensated from the smart contract.
The pseudocode of $\smartcontractinsurance$ is given in \cref{appendix:insurance smart contract}.

\parhead{Security analysis}
We briefly analyze how the initial goal was achieved with the design of the ``stick''.
For any attack, the service provider can earn at most the total value of all the accounts.
Therefore, as long as the collateral required to be put down is larger than this total amount,%
\footnote{We believe that a 100\% deposit is reasonable because the cost to the service provider is the potential interest they could have earned on the deposit, not the deposit itself.}
a service provider has no incentive to misbehave,
as each misbehavior costs more than what it gains. 

\section{The Carrot}
\label{sec:bounty}
In this section, we describe how we design the bounty (the carrot in \snc) to defend against the outsider attacker.
The goal is to encourage the outsider attacker to report the wallet signing key breach to the service provider without abusing the signing key.

Throughout this section, we refer to the service provider as the defender, using these two terms interchangeably.

\subsection{Desired properties of the Bounty}
\label{subsection: desired properties of bounty}
Distributing signing key shares across multiple TEEs with a threshold signature key generation procedure can lower the chance of signing key breaches caused by outsiders as used in \cite{herzberg1995proactive}.
However, it is not fully resolved.
In this section, we further mitigate the risk of unauthorized signatures resulting from side-channel attacks by external attackers with a bounty. 
The bounty enables the service provider to take appropriate actions before any catastrophic security breaches occur.

The two technical difficulties in the design of the bounty are: (1) how can the attacker and the service provider perform an atomic exchange of the key share and the reward; and (2) how to give the attacker just enough incentive to claim the bounty, while saving the defender's cost.
In detail, a good bounty should achieve the following goals:
\begin{enumerate} 
    \item An attacker gets the reward from the service provider if and only if he submits valid proof that convinces the service provider that he has obtained the key share.
    \item The construction itself does not leak any knowledge about the key share other than what has already been obtained by the attacker.
    \item An attacker prefers submitting the key share(s) to bounty over selling them in the market.
    \item An attacker submits the key share as soon as he gets the first key share, instead of continuing the attack.
    \item The defender's cost is minimized. 
\end{enumerate}

We suggest using smart contract bounty (\cref{subsection: the smart contract bounty}) to satisfy the goal 1-2.
The goals 3-5 are achieved by carefully designing a reward function for submitting key shares for a bounty claim.

\subsection{The Smart Contract Bounty}
\label{subsection: the smart contract bounty}
To realize the atomic exchange of the key share and the reward,
we initiate the bounty using a smart contract $\scbounty$.

As a defense against the outsider attacker,
the signing keys are rotated every $T$ units of time.
Following each key shares rotation, each TEE computes the hash of all the shares they hold and outputs the hash values to the service provider.
The service provider then publishes them in the $\scbounty$.
The problem arises when the service provider publishes the hash values that do not match the ones generated by the TEEs, 
making the bounty unable to be claimed.
To ensure that the hashes of the key shares are successfully published on the blockchain, we use the proof of publication scheme \cite{cheng2019ekiden}.
In other words, after each rotation or restart, the TEE will verify that the hash of the key shares they are using is the same as the latest version published on the blockchain (via proof of publication).
Only then will it use the current key shares to sign the user’s requests.

To claim the bounty,
the attacker submits the share(s) he finds as proof of knowledge.
To prevent front-running, proofs are submitted following a commit-and-reveal scheme \cite{commitreveal}.
We model this hash function as a random oracle so that it does not leak any information about the key shares themselves.

Upon receiving the key share, the smart contract $\scbounty$ checks whether the hash of the share is included in the smart contract.
If it is, $\scbounty$ puts the reward on hold for a designated period
and immediately invalidates all the current secret shares (such that the attacker cannot sell the shares or produce unauthorized signatures after submitting to the bounty).
At the same time, the service provider asks the user of the affected accounts to submit insurance in case there exists an unauthorized signature. 
The attacker gets the reward if there is no insurance claim for the signing key whose shares they are submitting.
The amount of the reward is determined by the reward function specified in \cref{subsection: reward function design}.
The formal protocol of bounty is given in \cref{appendix: bounty}.

\subsection{Reward Function Design}
\label{subsection: reward function design}

In this subsection, we apply a two-step methodology to the design of the reward function. 
First, we present a case study focused on the reward function for a deterministic attacker (\cref{subsubsection: Case Study for deterministic attacker}). 
Then, we broaden the scope to more general scenarios involving non-deterministic attacks (\cref{subsubsection: Proposed reward function for non-deterministic attacker} to \cref{subsubsection: Evaluation Results}), using observations and insights gained from the simpler case.

\subsubsection{Notation and Definition}
\label{sec:rwdfunccond}
In this section, we address two types of attackers:
the deterministic attacker and the non-deterministic attacker.
The deterministic attacker has a fixed deterministic cost function $C(k)$, which is analyzed in \cref{subsubsection: Case Study for deterministic attacker}.
The non-deterministic attacker has a fixed cost $\costPerStep$ of attacking one TEE at one step with a certain probability $\psuccess$ of obtaining one share of the key from the TEE at that step.
We deal with them in \cref{subsubsection: Proposed reward function for non-deterministic attacker} to \cref{subsubsection: Evaluation Results}.

In the smart contract bounty, the reward given to the attacker is determined by a reward function $R(k, t)$,
where $k$ is the number of shares that the proof is trying to prove against (i.e., the number of shares obtained by the attacker),
and $t$ is the submission time (which is the blockchain timestamp of the inclusion of the bounty claiming transaction).
Essentially, at time $t$, the attacker provides evidence of having acquired $k$ shares. 
Since the signing key is rotated every $T$ units of time and the signing key is secret-shared into $\keycap$ shares, we have $t \in [0, T]$ and $k \in [0, \keycap]$.

Recall that we use a $(m, \keycap)$ signature scheme. The service provider has $\keycap$ secrets shares, with $\geq m$ of them together having value $\amtcap$ 
for some $m \leq \keycap$,
and $k < m$ of them have value $\amtcap \cdot k/m$. \footnote{Note that in some cases, it may also make sense that having $k < m$ of them has no value.
For generality, we consider them to have some partial value.}
Since $m$ shares are enough to recover the key, the value of $m$ or more shares is the wallet value (i.e. $V(m) = V(m+1) = \cdots = V(N) = \amtcap$).
A notation table is provided in \cref{appendix: notation table}.

\subsubsection{Case study for deterministic attacker}
\label{subsubsection: Case Study for deterministic attacker}
We first provide a case study with respect to a simpler attacker: 
he has a deterministic cost function $C(k)$, which is non-decreasing in $k$, the number of acquired shares.  

\parhead{Naive solution} We start with a naive solution as briefly discussed in \cref{sec:intro}:
the linear reward function.
In other words, $R(k, t) = V(k) + \eta_{1}$ for some $\eta_{1} > 0$.
This is a natural solution:
it gives a bit more than how much the share(s) are worth.
However, as mentioned,
this naive solution can only achieve goal (3), but not (4) or (5) proposed in \cref{subsection: desired properties of bounty}.
As analyzed, the attacker would continue to attack for more shares and only submit when he has all the key shares.

\parhead{A starting point}
Therefore, we propose first a simple solution
that can achieve the goals 3-5 under such a deterministic attack (as the starting point for our real reward function):
\begin{equation*}
    R(k) = \max_{0 < k \leq \keycap}(V(k) - C(k)) + C(1) + \eta_{0} + (1 - t/T) \delta_{0},
\end{equation*}
where $\eta_{0}$ and $\delta_{0}$ are small constant numbers serving as bonus.
This reward function straightforwardly satisfies our goals. 
For goal (3): Submitting to the bounty provides the attacker with at least $\eta_{0}$ more than selling the shares when the attacker submits with only one share. Consequently, there is no incentive for the attacker to sell the share.
For goal (4): Since the adversary achieves maximum profit from the bounty by obtaining just one share $\max_{0 \leq k \leq \keycap}(V(k) - C(k)) + \eta_{0} + (1 - t/T) \delta_{0}$, and given that the bonus $\delta_{0}$ decreases over time, the attacker is incentivized to submit the share to the bounty upon acquiring the first share (and since the adversary needs one share to submit, $C(1)$ is used to compensate this cost).
For goal (5): the defender's cost is minimized since the defender cannot spend less. 
If she reduces her expenditure by $\eta_{0}$, the adversary's gain from the reward might equal the profit from selling the key at point $i$, where the profit $(V(k) - C(k))$ is maximized. This could lead the attacker to opt for selling the key.
As a side property, the attacker also saves cost,
as its total cost is always non-decreasing. 

\begin{figure}
    \centering
    \includegraphics[width=0.42\columnwidth]{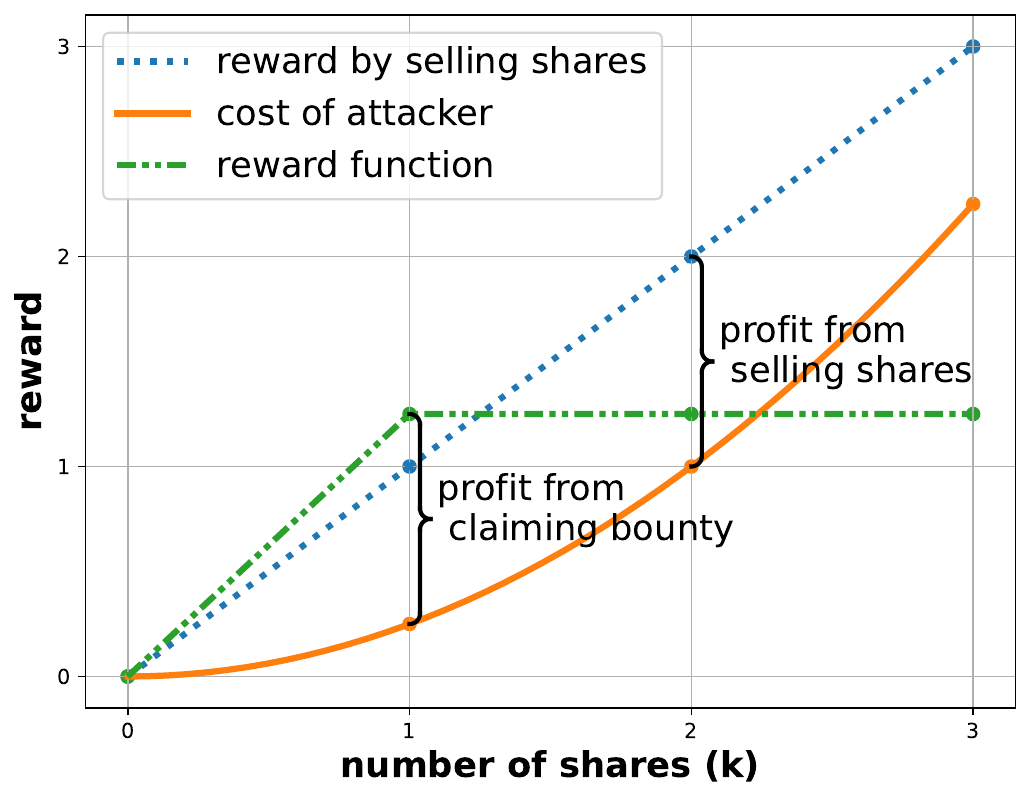}
    \caption{Example of reward function in simplified case.}
    \label{fig:intuition}
\end{figure}

A concrete example is depicted in \cref{fig:intuition}. 
Here, the cost of attack is $C(k) = \frac{1}{4} k^2$, and the value of key shares is $V(k) = k$. 
The maximum profit for the attacker is $\max_{0 \leq k \leq \keycap}(V(k) - C(k)) = V(2) - C(2) = 1$. 
We set $\eta_{0} = \delta_{0} = 0.1$.
Therefore, the optimal reward function in this scenario is $R(k) = C(1)+ (V(2) - C(2)) + \eta_{0} = 1.25 + \eta_{0}$. 
By structuring the reward function in this way, we not only incentivize the attacker to submit the key share as soon as they get one share but also reduce the defense cost.

Let's compare the reward function we proposed with two baselines: a zero function $R_0 (k) = 0$ and a linear reward function $R_l(k) = k + \eta_{0}$.
With $R_0$, the attacker accumulates 2 shares and sells them in the market, which violates goals 3 and 4.
With $R_l$, the defender pays $2 + \eta_{0}$ to prevent the attacker from selling 2 shares, which violates goal 3 and costs more than our reward function.

The main observation from the case study is that giving the attacker more reward at first share is not only a good way to persuade the attacker not to further exploit the key, 
but also saves the defender's cost.

Of course, here, the context is greatly simplified: the attacker's cost is a known deterministic function of the number of key shares gained.
If the attacker's cost is a probabilistic function, the reward function does not always achieve the goals.
Also, even for a deterministic attacker with a slightly different cost function, the reward function may not work anymore (e.g., if the attacker costs 10\% less per share).
Thus, we propose a more complete reward function in \cref{subsubsection: Proposed reward function for non-deterministic attacker}.

\subsubsection{Metrics for Reward Function}
\label{subsubsection: Metrics for Reward Function}

While for the deterministic attacker,
the simple reward function satisfies all the goals, it becomes more complicated for a non-deterministic attacker,
and also when we want to protect against a wider range of attackers.
There is a trade-off between goals 3-5 in $\cref{subsection: desired properties of bounty}$.
For example, it would cost more if we wanted to encourage the attacker to turn in the key shares to the bounty earlier.
To address this, we turn the goals into numerical metrics and balance them using a weighted average.

We developed three metrics to evaluate how well the reward function meets each of the three specified goals. 
The first metric is the probability of key shares being sold, denoted as $\probsellingkey$ (goal (3)). 
The second metric is the average holding time, $\holdingtime$, representing the average time between the attacker finding the first share and the termination of the game (goal (4)). 
The third metric, the cost to the defender, is denoted as $\costofdefender$ (goal (5)). 
The cost of the defender is the max between the value the attacker gets by selling the $k$ shares (i.e., $V(k)$) and the amount of the bounty claimed (recall that an attacker can only do one of the two instead of both). 
To combine these metrics into a score, denoted as $\fscore$, we introduce parameters $\alpha_1$ and $\alpha_2$ to compute a weighted average.
\begin{equation}
\label{eq: fscore}
    \fscore = \alpha_1 \cdot \probsellingkey + \alpha_2 \cdot  \frac{\holdingtime}{\totaltimesteps} + (1 - \alpha_1 - \alpha_2) \cdot  \frac{\costofdefender}{\amtcap}
\end{equation}

In \cref{eq: fscore}, the holding time is normalized by the time period $T$ and the defender's cost is normalized by the value of the key $\amtcap$.  %

\subsubsection{Propose reward function for non-deterministic attacker}
\label{subsubsection: Proposed reward function for non-deterministic attacker}

We now propose a reward function designed to achieve the objectives outlined in \cref{subsection: desired properties of bounty} for a non-deterministic attacker. 
The optimization and evaluation of this proposed reward function will be detailed in the subsequent parts of this subsection.

To achieve goal (3) in \cref{subsection: desired properties of bounty}, we need to give more reward to the attacker than the value of the shares. 
For an attacker with $k$ shares of secret,
he can gain $V(k)$ units of money.
Thus, to encourage the attacker to submit to the bounty,
we give out more than the amount they should have received by selling the key shares.
A non-deterministic attacker, however,  may get lucky in some cases and get more than one share at a low cost. 
So our proposed function should have the property $R(k, t) > V(k)$ for all $k \in [1, N]$.

Formally, we give a reward of $V(\keycap)^\epsilon\cdot V(k)^{1 - \epsilon} + \eta$ (recall that $dV/dk \geq 0$ for all $k \in [\keycap])$,
for some $\epsilon \in [0, 1], \eta > 0$.
As long as $\epsilon \geq 0, \eta > 0$,
we have $V(\keycap)^\epsilon\cdot V(k)^{1 - \epsilon} + \eta > V(k)$ for all $k > 0$.
Note that when $\epsilon$ increases, we give more reward when $k = 1$, which could potentially reduce the defender's cost (achieving the goal (5)) according to the case study above.

Finally, we need to encourage the adversaries to submit earlier to achieve goal (4) in \cref{subsection: desired properties of bounty}.
Similarly, we set the ``extra bonus'' decreasing overtime.
Formally,
let $g(k) := V(\keycap)^\epsilon\cdot V(k)^{1 - \epsilon} + \eta - V(k)$ denoting the extra reward we paid to the attacker.
We reduce this gain by time: adding a term $-g(k) \cdot t/T$. 
The reward function we suggest is:
\begin{equation}
\label{eq:reward}
    R(k,t) := V(\keycap)^\epsilon\cdot V(k)^{1 - \epsilon} + \eta - g(k) \cdot t/T,
\end{equation}
where $g(k) := h(k) + \eta - V(k)$.
$\delta \geq 0$, and $\eta > 0$.

To model the real-world constraint of the defender's budget, we also introduce an additional parameter, $\bountycap$, into the reward function.
This parameter represents the maximum amount of money that the bounty can afford, expressed as a percentage of the secret's value.
Specifically, we add a bound $\bountycap \cdot V(\keycap)$ to our reward function $R(k, t)$ (\cref{eq:reward}),
and
the resulting new reward function is:
\begin{equation}
    \label{eq: bounty cap}
    \tilde{R}(k, t) = 
    \begin{cases}
        R(k, t) & \text{if $R(k, t) < \bountycap \cdot V(\keycap)$ }\\
        \bountycap \cdot V(\keycap) & \text{if $R(k, t) \geq \bountycap \cdot V(\keycap)$}
    \end{cases}
\end{equation}
where $t$ is the submission time and $k$ is the number of submitted shares ($t \in [0, T], k \in [0, \keycap]$).

\subsubsection{Modelling the non-deterministic attacker}
\label{subsubsection: Modelling the non-deterministic attacker}
To evaluate our function,
we first need to model how an attacker behaves.
To do this, we first describe the behavior of the attacker that can be modeled as the optimal stopping game.
Then,
we further find the optimal attacker strategy using a Markov decision process (MDP).

Moreover, with this evaluation result,
the defender can quantitatively understand what range of attackers can be effectively prevented using this reward function.
She can then change the parameters (e.g., the attacker's ability to begin with and the budget) to modify the function accordingly.

\parhead{Attacker behavior}
We give a detailed description of the attacker's decision process as follows. 
As in the preceding sections, we exclusively consider a single signing key that is shared among $\keycap$ TEEs.
The time period during which the secret remains valid is divided into $\totaltimesteps$ discrete time steps. 
Each time step is further divided into two sub-steps, during which the attacker makes distinct choices:
In the first sub-step, the attacker selects the number of TEEs to target during that step.
In the second sub-step, the attacker decides whether to terminate the game (sell the shares or claim the bounty) or proceed to the next step.
If an attacker decides to target a TEE in a given step, they have a success probability of $\psuccess$ to acquire a key share from it, while incurring a fixed cost of $\costPerStep$.

\parhead{Optimal stopping game}
We model an adversary as a player of an ``optimal stopping'' game \cite{optimalstopping1,optimalstopping2,optimalstopping3}.
Essentially,
the optimal stopping game states the following:
there is a sequence of random variables $X_1, X_2, \dots$ whose distribution is assumed to be known;
and there is a sequence of gain functions $(Y_i)_{i \geq 1}$ which take the first $i$ random variables as inputs (i.e., $Y_i(x_1, \dots, x_i)$ is a function over $x_1 \gets X_1, \dots, x_i \gets X_i$). 
Then, the player observes the sequence of random variables one at a time,
and for each step $i$, the player can either stop observing and claim the gain $Y_i(x_1, \dots, x_i)$ or continue.
The goal of the player is to optimize the expected gain.
Note that this setting is essentially the same as our setting, where the random variables are the shares gotten by the adversary (e.g. if an attacker can obtain a share with probability $p$ at step $i$, $X_i$ is a Bernoulli random variable returning $1$ with probability $p$ and $0$ with probability $1-p$).
Then, $y_i$ is the profit the attacker can gain from all the shares he has obtained up to step $i$,
which is the maximum between the value of the bounty and the value of selling these shares, less his cost up to step $i$.
Although some specific forms of optimal stopping games have closed-form solutions (e.g., the secretary problem \cite{secproblem}), for more complex scenarios like ours, a typical approach to find the player's optimal strategy is to model the game with Markov Decision Process (MDP) \cite{optimalstopping1,optimalstopping2}.

\parhead{MDP}
We model the attacking process as an MDP,
structuring it into discrete steps. 
At each step, the attacker decides the number of TEEs    to target. 
The attacker also needs to determine the optimal time to end the attack and obtain their reward: after each step, he must choose to either cease the attack and get the reward or continue attacking in the subsequent step.

We specify the state transition function and the reward function of the MDP as follows.
The state of the MDP is defined by the tuple of the number $k$ of shares gained by the attacker, the time slot $t$, and the sub-step in each time slot $d \in \{0, 1\}$.
At state $(t, k, 0)$, the attacker needs to choose the number of TEEs (denoted as $n$) to attack in this time slot.
The state transitions to $(t, k + \Delta k, 1)$, where $\Delta k$ is the number of key shares gained in this time slot. 
The number of newly gained key shares depends on the success rate $\psuccess$ and the number of TEEs the attacker chooses to attack in that particular step.
Specifically, the probability that the attacker gets $i$ new shares in this time slot is $Pr(\Delta k = i) = \binom{n'}{i} \psuccess^{n'} (1 - \psuccess)^{n'-k}$, where $n' = \max (n, m-k)$.
At state $(t, k, 1)$, the attacker faces a decision: either end the game by selling the key shares or submitting them to the bounty, or wait until the next time slot.
If the attacker chooses to wait until the next time slot, 
the state will transition to state $(t+1, k, 1)$.
If the attacker chooses to sell the key shares or submit them to the bounty,
the next state will be the termination state.
When the time slot reaches the maximum time $T$ at state $(t, T, 1)$, the next state will be the termination state.

At each step of the process, the attacker incurs a negative reward of $-\costPerStep \cdot n$, representing the cost of the attacking $n$ TEEs.
The attacker gains a positive reward $R(k,t)$ if he submits the key shares to the bounty.
Alternatively, if he decides to sell the key shares, 
he gets $V(k)$.
A summary of the transition and reward function of the decision problem is in \cref{table: MDP transition table}.

\begin{table}[htbp]
\caption{Description of the state transition and reward matrix}
\centering
\begin{tabular}{|l|l|l|l|}
\hline
State $\times$ Action & State & Probability & Reward \\ \hline
$(k, t, 0) \times $attack $n$ TEEs   & $(k+i, t, 1)$   &  $Pr(\Delta k = i)$  & $-n \cdot \costPerStep$   \\ \hline
$(k, t, 1) \times$ wait $(t<T)$   & $(k, t+1, 0)$   & 1   & 0   \\ \hline
$(k, t, 1) \times$ wait $(t=T)$   & termination   & 1   & 0   \\ \hline
$(k, t, 1) \times$ turn in  & termination  & 1  & $R(k, t)$  \\ \hline
$(k, t, 1) \times$ selling key   & termination  & 1  & $V(k)$  \\ \hline
\end{tabular}
\label{table: MDP transition table}
\end{table}

Utilizing the MDP solver \cite{mdptoolbox}, we are able to compute the attacker's optimal strategy for a specific reward function. By examining this optimal strategy, we can obtain the metrics defined in \cref{subsubsection: Metrics for Reward Function} ($\fscore$ score). The $\fscore$ score then serve as the objective for optimizing the parameters within the reward function.

\subsubsection{Optimize the Reward Function Parameter}
\label{subsubsection: Optimize the Reward Function Parameter}

In this part, we describe the methodology for deciding the optimal $\epsilon$ within the reward function in \cref{eq:reward}, with $\bountycap$ as described in \cref{eq: bounty cap}. 

Recall that our reward function $\tilde{R}$ is determined by $\bountycap$ (bounty cap), $\epsilon$ (determining the starting point of the reward), and $\delta$ (how fast the reward decays by time).
We assume $\bountycap$ is some constant predefined by the defender, according to her budget.

We now explain our approach for identifying the optimal value of $\epsilon$ with regard to the performance metric $\fscore$.
As the defender aims to minimize the cost of the defender, the probability that the attacker will sell the key on the market, and the holding time, the objective is to minimize the score $\fscore$.
When defending against an attacker, the service provider must first decide the parameters used in $\fscore$ ($\alpha_1$ and $\alpha_2$) and estimate the ability of the attacker by specifying $\psuccess$ and $\costPerStep$.
Using the estimated parameters 
, an optimal $\epsilon$ could be numerically computed.
Specifically, we discretize $[0, 1]$ into a sequence of evenly spaced numbers, calculate a score for each $\epsilon$, and select the one corresponding to the lowest score.
\footnote{The precision is affected by how many intervals $[0,1]$ is discretized into.}

Upon determining the optimal $\epsilon$ with estimated parameters, we examine how attackers of various abilities respond to the computed $\epsilon$ in the next part. Specifically, these attackers might have different $\psuccess, \costPerStep$ compared to the initial estimates used for $\epsilon$ optimization, representing a range of adversaries stronger or weaker than the initial expectation.

\subsubsection{Evaluation Results}
\label{subsubsection: Evaluation Results}
We compare the score $\fscore$ of different reward functions, including our reward function, the linear reward function (see below), and no bounty (reward function equals 0).

The linear reward function is a solution that satisfies goal 3 without considering the cost.
Recall that we introduced this naive solution in \cref{sec:intro} and \cref{subsubsection: Case Study for deterministic attacker}:
in the linear reward function,%
the bounty claimer gets the exact value of share(s) plus a small bonus $\eta_{1}$ to encourage turning in key share(s).
We additionally set a time bonus $\delta_{1}$ that decays with time and encourages early turn-in for the purpose of this case study (to break ties for attacker decisions in MDP),
formally given as follows:
$
R_{l}(k,t) = V(k) + (1 - t/T) \delta_{1}  + \eta_{1}
$.

\begin{figure*}[htbp]
    \centering
    \includegraphics[width=\columnwidth]{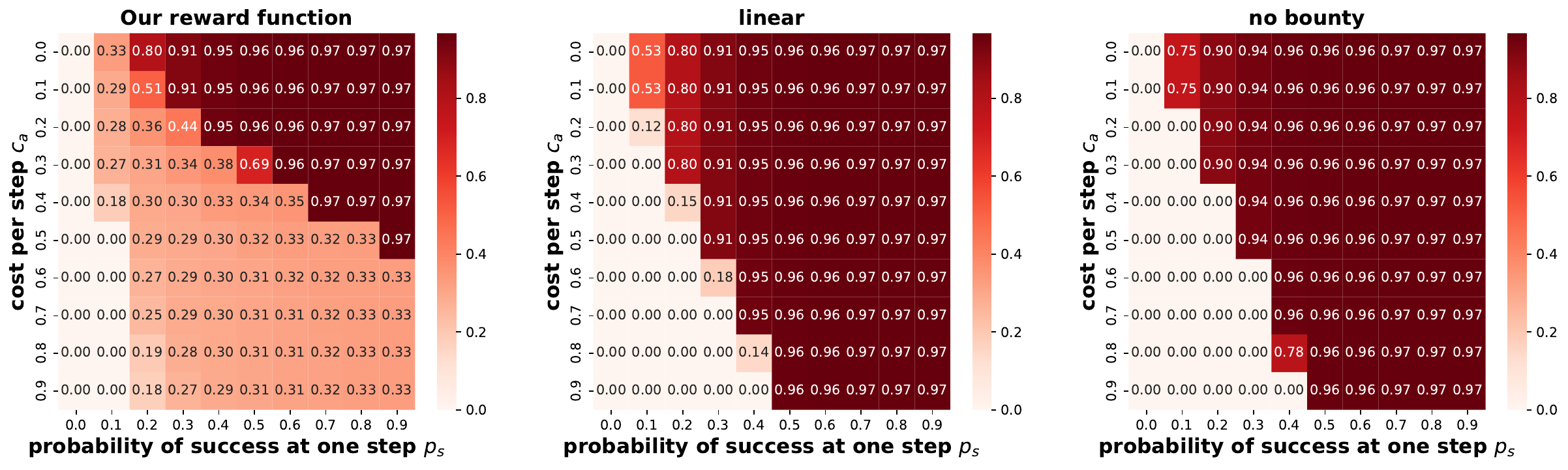}
    \caption{f score for different reward functions. $\bountycap= 0.8$. $\alpha_1 = \alpha_2 = 1/3$, $\costPerStep = 0.4$, $\psuccess = 0.4$, $\keycap = 3$, $\amtcap = 6$. Optimal $\epsilon = 0.95$. 
    }
    \label{fig:comopare reward functions}
\end{figure*}

In the evaluation, we set the estimation as $\costPerStep = 0.4$ and $\psuccess = 0.4$.
We set the total number of key shares as $\keycap = 3$ and the value of the key as $\amtcap = 6$ , which means the value per share is 2. 
In expectation, the cost incurred by the attacker to obtain one share is 1 (cost per step / probability of success), resulting in a positive expected profit of 1 for each share acquired .
We set $\alpha_1 = \alpha_2 = 1/3$ which means each metric has equal importance.
The parameters can be replaced with real-world values when the wallet is implemented in practice.
The optimal $\epsilon$ we get is 0.95 given the parameters above.
Then, we use the optimal parameter to derive the score for attackers with variant cost $\costPerStep$ and success rate $\psuccess$.   %

We show how this function behaves when facing to different attackers in \cref{fig:comopare reward functions},
where each cell within the heatmap shows the $\fscore$ score corresponding to a specific configuration of the attacker's capabilities, denoted by the parameters $\costPerStep$ and $\psuccess$.
When the cost is low and the success rate is high (located in the upper right region of the heatmap), the attacker is considered strong. 
Conversely, when the cost is high and the success rate is low (positioned in the lower left area of the heatmap), the attacker is perceived as weak.

As we can see in the heatmap, when $\bountycap = 80\%$, the performance of the reward function we proposed (state of the art) is better than the baseline (no bounty and linear reward function) in most cases.
For most attackers, regardless of the ability, our reward function generates a smaller score. 
The figure demonstrates that our reward function has great performance not just for attackers whose abilities are equal to our estimations ($\costPerStep = 0.4$ and $\psuccess = 0.4$), but it also works well for stronger attackers. 
As shown in the figure, essentially for \textit{any} $\psuccess$, as long as $\costPerStep \geq 0.4$,
the $\fscore$ score is at most $0.3$.
Similar flexibility on $\costPerStep$ can also be seen in the graph.
These results indicate that even without precise attacker ability estimations, our reward function outperforms the alternative reward functions
and shows decent effectiveness in preventing outsider attacks.

As mentioned,
the defender can then use the heatmap to determine the effectiveness of the reward function given the current attacker's ability estimation and the budget.
She may increase her budget to find a reward function that effectively defends against a broader spectrum if needed.

\section{Case Study}
\label{sec:case study}

We briefly discuss how to choose the parameters for the bounty in \snc using a simple case study.
Recall that we need to set time $T$,
the expected return given the number of shares $V(k)$,
and the cost function $C(k)$.
The calculation below assumes using a (10,20)-threshold signature scheme (i.e., 10 shares are enough to recover the secret) and $T = 30$.

To set the rest of the parameters,
we first examine the state-of-the-art side-channel attacks against ECDSA.
ECDSA \cite{ecdsa} is the most commonly used signature scheme for blockchains like Bitcoin \cite{bitcoin},
and thus we use it as an example.
To our knowledge, all the side-channel attacks without root privilege in recent years against the most popular ECDSA library
(OpenSSL \cite{openssl}) show that they require at least $2^{12}$ traces
to recover a secret \cite{ecdsaattack1,ecdsaattack2,ecdsaattack3}.
Then, we let the service provider cap the number of signatures a user can make.
According to \cite{atlantafed2020},
a regular user makes 68 bank transactions per month,
which means $\sim 2.3$ transactions per day.
To be lenient, assume the victim makes 230 transactions per day (which is 100x the average number of transactions per day).
Since recovering a key share requires at least $2^{12}$ signatures, which takes $\sim 17.8$ days. 
For $V(k)$,
recall that we have a rate limit $\amtcap$ for each wallet (i.e., the amount of money in each wallet).
According to \cite{federal},
each transaction's average value is 36 dollars for a debit card.
We thus set $\amtcap = 36000$, again 100x larger than the average transaction value.
Each key share has equal value, and $\thresholdmat = 10$ shares are enough to recover a key,
we set $V(k) = \min(\ceil{\amtcap \cdot k/\thresholdmat}, \amtcap)$.

Lastly, we discuss the cost function.
The cost function is the most tricky one,
since it should capture all the possible costs of an attacker,
including operational costs, the risk of being caught,
the side channel being mitigated, and so on.
Thus, we propose a conservative function (i.e., the minimum cost an attacker can have).
Note that for an outsider, the minimum requirement is essentially
getting to obtain the traces remotely.
The most common way is residing on the same virtual machine as the victim program, as discussed in \cite{RTSS09}.
Thus, we estimate the cost using the cost of renting the same cloud machine as the service provider.
Suppose that it costs $\rentingamt$ dollars per unit of time (e.g., c5.metal from AWS, a commonly used server instance, costs ${\sim}\$97.9$ per day \cite{awsprice}).
Thus, we have $C(k) = \rentingamt \cdot k \cdot 17.8$.

These numbers give us that to recover a key with a value of $36000$ dollars,
the cost of the attacker is at least $\sim 17426$ dollars (based on 17.8 days per share, a total of 10 shares, and 97.9 dollars per day for VM).
We can come up with a reward function accordingly given all these numbers, along with their budget limit.
More accurate numbers can be obtained for a specific service provider by analyzing their own transaction data.

\section{Discussion}
\label{section: discussion}
In this section, we discuss \snc's performance, limitations and extension application.

\thenewparagraph{Performance Analysis}
Reasonable signing performance is required to make the scheme practical.
A potential bottleneck of performance may be caused by the secret sharing between different TEEs. 
In this part, we analyze its concrete performance to show that the multi-TEE ECDSA signing will not be a bottleneck.

For the threshold ECDSA scheme proposed by Gennaro and Goldfeder~\cite{gennaro2018fast},\footnote{This scheme considers malicious participants, so there are unnecessary steps in the protocol if we assume all the participants are honest, which is true in our case.}
the benchmark for the signature generation time among $m$ participants is $29 + 24m$ milliseconds.
As benchmarked in \cite{mofrad2018comparison}, 
the highest overhead of TEE is $19.31\times$ in all the tasks tested.
Therefore, a conservative signature generation time is around $560 + 463m$ milliseconds. 
The protocol requires five rounds of communication and we estimate the communication delay for each round as $100$ milliseconds~\cite{de2008one}.
Consequently, the total time for generating a threshold signature is about $1060 + 463m$ milliseconds,
which is generally acceptable for cryptocurrency wallets.
Additionally, to accommodate high transaction volumes, we can employ multiple sets of TEEs in parallel.

\thenewparagraph{Limitations of insurance}
Our techniques provide a technical basis for penalizing the service provider when an attack succeeds against it, providing an incentive for it to properly safeguard its TEEs from outside attackers and a transparent and measurable guarantee to end users.  
These are significant improvements over the current status quo.  
Ensuring that the company deposits assets sufficient to satisfy claims against it is a matter for insurance regulators; today, insurance regulators in most jurisdictions require companies to maintain \textit{statutory reserves}, i.e., an amount of cash and readily marketable securities that it can use to pay its foreseeable claims.  
As with other insurance in real life (e.g., property insurance), users in our system may not be compensated if these reserves (i.e., the company's deposits) are depleted by other claims.  
Our technical solutions presented here cannot entirely eliminate the need for legal recourse in such situations.  Nevertheless, our design provides a stronger foundation for reducing trust in a service provider and for reducing the risk of clients.

\thenewparagraph{Limitations on the type of assets}
Note that in most blockchains today,
each wallet is tied to a specific private key.
Thus, key updates after leakage can cause the assets in the wallet to be non-retrievable.
In our paper,
we require the asset to be tied to a smart-contract-based wallet, allowing the key updates to work as expected.
How to extend our idea to support a wallet without such support remains open.

\parhead{Another application: CA}
To further demonstrate the flexibility of our design, we apply 
the \snc (with modification) to the Certificate Authority (CA) system~\cref{sec:ca}.

\section{Conclusion}

In this paper, we introduced \snc, a solution designed to mitigate side channels in TEE-based cryptocurrency wallets by leveraging economic incentives. 
Our wallet authentication system utilizes OAuth to ensure both accountability and user-friendliness. Additionally, we designed a combination of stick (insurance) and carrot (bounty) to safeguard against both insider and outsider attacks. Finally, we evaluated our approach and showed its effectiveness.

\bibliography{reference}

\appendix
\section{Preliminaries}
\label{sec:formalprimitive}

We present the formal definitions of the cryptography primitives.

\parhead{Digital signature scheme}
A signature scheme has the following three PPT algorithms:
\begin{enumerate} 
    \item $(\vk, \sgk) \gets \keygen(1^\lambda):$ takes a security parameter $\lambda$, and outputs a verification key $\vk$ and a signing key $\sgk$.
    \item $m_\sigma \gets \sign(\sgk, m):$ signs a message $m \in \{0,1\}^*$ with signing key $\sgk$, and outputs a signed message $m_\sigma$.
    \item $b \gets \verify(\vk, m_\sigma):$ verifies a a signed message $m_\sigma \in \{0,1\}^*$ against a verification key $\vk$.
\end{enumerate}
We adopt the standard correctness and non-forgeability definitions.

\parhead{Threshold digital signature scheme}
An $(n,k)$-signature scheme has the following three PPT algorithms:
\begin{enumerate} 
    \item $(\vk, \sgk_1, \dots, \sgk_n) \gets \keygen(1^\lambda):$ takes a security parameter $\lambda$, and outputs a verification key $\vk$ and $n$ shares of signing key $\sgk_1, \dots, \sgk_n$.
    \item $m_\sigma \gets \tsign(\sgk_1, \dots, \sgk_k, m):$ signs a message $m \in \{0,1\}^*$ with signing key $\sgk$, and outputs a signed message $m_\sigma$.
    \item $b \gets \verify(\vk, m_\sigma):$ verifies a a signed message $m_\sigma \in \{0,1\}^*$ against a verification key $\vk$.
\end{enumerate}
We adopt the standard correctness and non-forgeability definitions.

\parhead{Public Key Encryption (PKE)}
A PKE scheme has the following three PPT algorithms:
\begin{enumerate} 
    \item $(\pk, \sk) \gets \keygen(1^\lambda)$: takes a security parameter $\lambda$ as input and and generates  a key pair $(\pk,\sk)$.
    \item $\ct \gets \Enc(\pk, m)$: encrypts a message $m \in \{0,1\}^*$ under a public key $\pk$.
    \item $m' \gets \Dec(\sk, \ct)$: decrypts a ciphertext $\ct$ with a signing key $\sk$.
\end{enumerate}
We adopt the standard correctness and semantic security definition for PKE.

\section{Notation Table}
\label{appendix: notation table}

A notation table of the symbols used for the reward function design (\cref{sec:bounty}) is given in \cref{table: notation}.

\begin{table}[htbp]
\centering
\caption{Notation Table}
\label{table: notation}
\begin{tabular}{|c|l|}
\hline
Symbol & Description \\ \hline
$m$    & threshold in threshold signing scheme \\ \hline
$N$    & total number of key shares \\ \hline
$k$    & number of key shares gained by the attacker \\ \hline
$\totaltimesteps$  & total number of time steps in MDP model \\ \hline
$V(k)$    & value of $k$ key shares \\ \hline
$\amtcap$ & value of the full key \\ \hline
$\probsellingkey$ & probability that the attacker sells the key shares \\ \hline
$\holdingtime$ & average time that the attacker holds the first share \\ \hline
$\costofdefender$ & cost of defender \\ \hline
$\alpha_1, \alpha_2$ & weight parameter in the metric function \\ \hline
$\fscore$ & metric for the reward function \\ \hline
$C(k)$ & cost function of the deterministic attacker \\ \hline
$\costPerStep$ & cost of per step \\ \hline
$\psuccess$ & success rate per step \\ \hline
$R(k, t)$ & reward function of the bounty \\ \hline
$\epsilon$ & shape parameter of the reward function \\ \hline
$\eta$ & bonus in the reward function \\ \hline

\end{tabular}
\end{table}

\section{Details of Wallet and Insurance Smart Contracts}
\label{appendix: details of insurance and wallet}
In this section, we give more details about the workflow of the wallet, focusing on transaction signing request and insurance claim.
In our paper, we name the smart contract based on their functionality (e.g. $\scwallet$ for transaction execution).
However, in the real implementation, the functions can be integrated in one smart contract.

\subsection{Transaction request through smart contract}
\label{appendix:TEE Availability}
\begin{figure}[htbp]
  \centering
  \includegraphics[page=3,width=0.6\columnwidth]{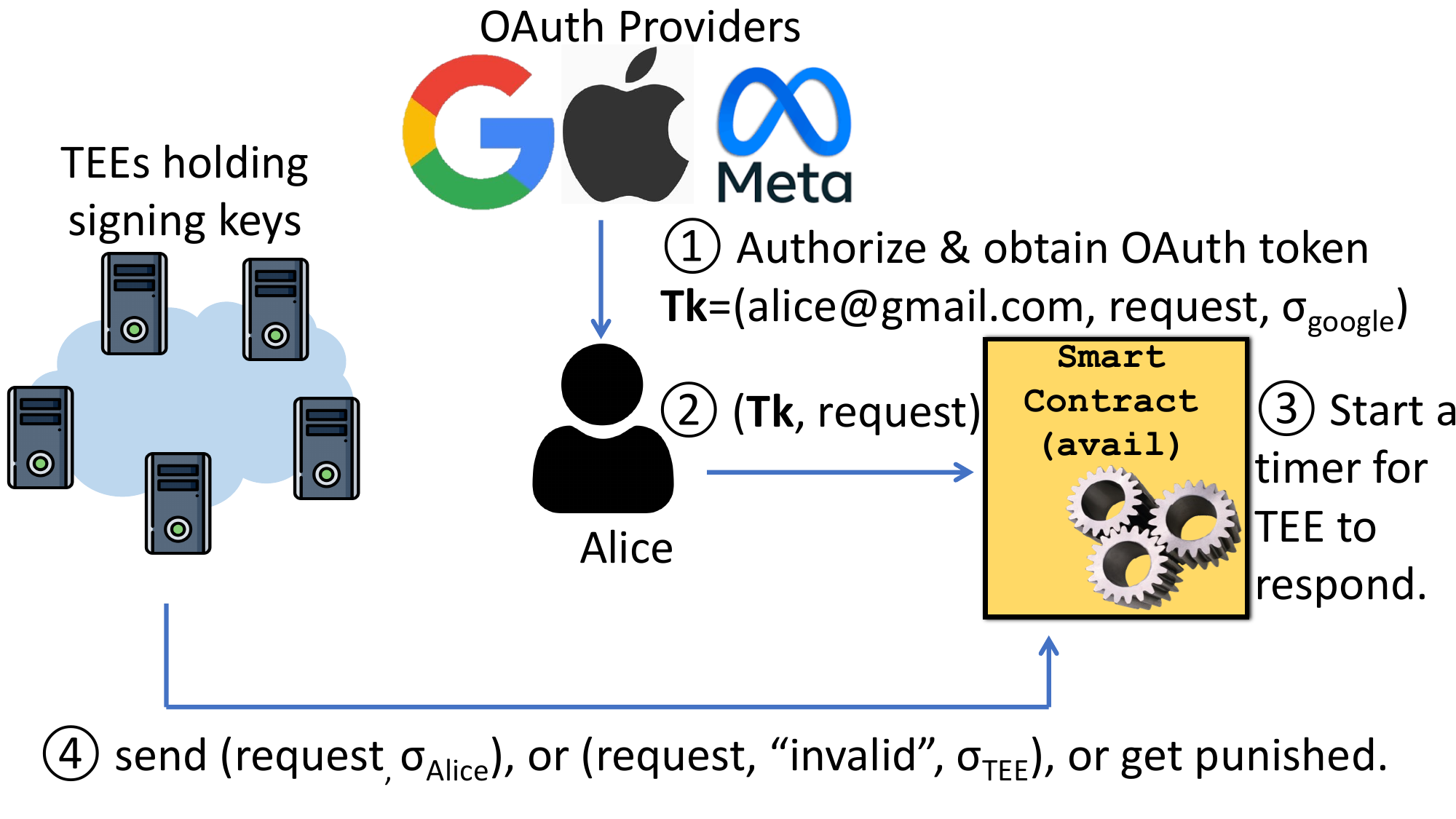}
\caption{Signing request via $\smartcontractTEESC$.}
  \label{fig:overview_framework3}
\end{figure}

$\smartcontractTEESC$ is used to ensure that the TEE respond to user's signing request in time.
The workflow of $\smartcontractTEESC$ is given in \cref{fig:overview_framework3}.
Once requested by the user, the service provider must provide the signature or a message signed with the TEE within a predefined time; otherwise, the deposit in the smart contract will be burnt.

\subsection{Shutdown and Recovery}
After a valid insurance or bounty claim, the wallet service shuts down (by raising the global red flag in the insurance or bounty smart contract) and all transactions cease.
The service provider then generates new key pairs (again via a threshold signature scheme using TEE) for all the users to make sure the original ones obtained by the attacker do not work anymore. 
The balance of the old wallets will be transferred to the new smart contract wallets, replacing the old wallet keys with the new ones.
After the key-regeneration and replacement, the system restores.
The TEE will sign a message with its attestation key that will clear the red flag in $\smartcontractinsurance$ or $\scbounty$.

\subsection{Wallet Smart Contract}
\label{appendix:wallet smart contract}
We create a wallet smart contract $\scwallet$ for each of the user.
There is a global red flag which can be triggered when the insurance or bounty is claimed.
No token can be transferred out of the contract when the red flag is on, unless the transaction is signed by the TEE attestation key ($\pkattestation$).

If the user wants to transfer the money out, $\scwallet$ first checks that the global red flag is off, and then verify the signature.
If the account has enough balance, the balance will be deducted and the amount of token specified in the transaction will be transferred to the specified destination.

\begin{algorithm}
\caption{Wallet}
\label{algo:wallet}
\begin{algorithmic}[1]
\Function{Transfer}{tx, signature}
\If{isRedFlagOn == True}
\State \Return ``Red Flag is on''
\EndIf
\State (fromAddr, toAddr, amount) $\leftarrow$ tx
\If{verifySignature(tx, signature, fromAddr) == False}
    \State \Return ``Invalid signature''
\EndIf
\If{balance < amount}
    \State \Return ``Insufficient balance''
\EndIf
\State balance -= amount
\State \Call{transferToken}{toAddr, amount}
\State \Return ``Transfer Success''
\EndFunction

\Function{ReplaceKey}{tx, signature}
\State (fromAddr, toAddr) $\leftarrow$ tx
\If{verifySignature(tx, signature, $\pkattestation$) == False}
    \State \Return ``Invalid signature''
\EndIf
\State balance $\leftarrow$ 0
\State \Call{transferToken}{toAddr, balance}
\State \Return ``Key Replacement Success''
\EndFunction

\end{algorithmic}
\end{algorithm}

\subsection{Insurance Smart Contract}
\label{appendix:insurance smart contract}
The details of the workflow for an insurance claim is given in \cref{algo:insurance}.
As shown in function CLAIM, to dispute a transaction, the user first request a token to prove her identity and then construct the message to be processed with the TEE.
After the message is submitted to the insurance smart contract (function INSURANCE), the service provider will feed the message to function REPLYCLAIM and send the response to the insurance smart contract.

\begin{algorithm}
\caption{Insurance}
\label{algo:insurance}
\begin{algorithmic}[1]
\Function{Claim}{$\txinfo, \compensation, \providerlist$} \Comment{User}
 \State $\disputestring \gets $"dispute"$||h(\txinfo)||\compensation$
 \State $\dtokens \gets \rqtokens$ ($\idlist, \providerlist, \disputestring$)
 \State $\query \gets \Enc([$``dispute"$, \dtokens, \txinfo, \compensation]$, $\pkenclave$)
 \State \Return $\query$ 
 \EndFunction
\end{algorithmic}

\begin{algorithmic}[1]
\Function{Insurance}{$\query$, tx} \Comment{Smart Contract}

\If{redFlag == 1} 
\State terminate
\EndIf
\If{\Call{isExecuted}{tx} == 0 or IsCompensated[tx] == 1} 
\State terminate
\EndIf
\State Store $\query$ on the blockchain and start the timer
\If{($\result, signature) \leftarrow \Call{ReplyClaim}{\query}$ is submitted in time and signature verifies}
\If{\result[0] = ``dispute is correct"} 
\State parse $\txinfo$ and $\compensation$ from \result[1]
\State Send $\txinfo.\txvalue + \compensationvalue$ units of money to $\compensation$ \Comment{Compensate the user}
\State IsCompensated[tx] = 1  \Comment{Mark the tx as compensated}
\State redFlag $\leftarrow 1$ \Comment{Raise the red flag}
\Else
\State Send ``Incorrect dispute'' to the user and terminate
\EndIf
\Else
    \State burn the deposit
\EndIf
\EndFunction
\end{algorithmic}

\begin{algorithmic}[1]
 \Function{ReplyClaim}{$\query$} \Comment{TEE}
 \State $\dtokens, \txinfo, \compensation \gets \Dec(\query)$
 \State $\idlist' \gets$ $\idmap$[$\txinfo.\source$].$\idlist$;
 \State $\disputestring \gets $``dispute"$||h(\txinfo)||\compensation$
\If{$\vrftokens$ ($\idlist'$, $\dtokens$, $\disputestring$) == 1}
\State $\txisrecorded \gets \txinfo \in \tkdb ? 1 : 0$
\If{$\txisrecorded = 0$}
\State $ \result \leftarrow [\textrm{``dispute is correct"}, \txinfo||\compensation]$
\Else 
\State $r \leftarrow [\textrm{``dispute is incorrect"}, \txinfo||\compensation]$
\EndIf
\EndIf
\State \Return $\result$, signature($\result$, $\skenclave$)
\EndFunction
\end{algorithmic}

\end{algorithm}

\section{Bounty}
\label{appendix: bounty}

\subsection{Proof of knowledge}\label{apd:pok}
In this section we give the formal definition of our PoK primitive in \cref{sec:bounty} and the pseudocode of our constructions.

\parhead{Definition}
\newcommand{\proofofknowledgedef}{
We start by addressing how the attacker can convince the verifier that he has obtained some secret shares.
To achieve this,
we require an additional building block: a proof of knowledge scheme (PoK),
containing three interfaces:
(1) $\Setup$ taking a security parameter $\lambda$ and a secret set $\sset$ and generates a public parameter $\pp$;
(2) $\ProofGen$ taking a subset of $\sset$ and $\pp$ and generates a proof $\pi$;
(3) $\verify$ taking a proof $\pi$ and $\pp$ and outputs 0 or 1.

The attacker can use it to prove that he has obtained the secret shares of the service provider (correctness).
Then, if the attacker provides proof,
he must have known the secret shares (extractability).
Moreover, this scheme should not leak information about the secret except for what is already learned by the attacker during the attack (security).

We formally define it as follows.

\thenewparagraph{Proof of knowledge.}
A proof of knowledge $\pok$ scheme has three PPT algorithms:
\begin{enumerate} 
    \item $\pp \gets \Setup(1^\lambda, \sset):$ takes a security parameter $\lambda$ and a secret set $\sset$, and output a public parameter $\pp$.
    \item $(\pi, k) \gets \ProofGen(\pp, \sset'):$ takes a public parameter $\pp$ and a secret set $\sset'$, and outputs a proof $\pi$ and the number of secrets $k$.
    \item $b \gets \verify(\pp, \pi, k):$ takes a public parameter $\pp$, a proof $\pi$ and the number of secrets $k$, and outputs a bit $b$.
\end{enumerate}
It also satisfies the following property:
\begin{enumerate} 
    \item (Correctness) If $\sset' \subseteq \sset$, let $\pp \gets \Setup(1^\lambda, \sset)$, $(\pi, k) \gets \ProofGen(\pp)$, $b \gets \verify(\sk, \pi, k)$, then $\Pr[b = 1] \geq 1 - \negl(\lambda)$.
    \item (Extractability) For any PPT adversary $\adv$ there exists some PPT extractor $\ext$ such that the following holds: let $\pp \gets \Setup(1^\lambda, \sset)$, $(\pi, k) \gets \adv(\pp)$, $b \gets \verify(\sk, \pi, k)$,
    and $b = 1$, then let $\sset' \gets \ext(\adv, \pp, \pi, k)$, $\Pr[(\sset' \subseteq \sset) \wedge (|\sset'| = k)] \geq 1 - \negl(\lambda)$.
    \item (Security) For any PPT adversary $\adv$, 
    for any two secret sets $\sset \neq \sset'$,
    let $\pp \gets \Setup(1^\lambda, \sset)$, 
    and $\pp' \gets \Setup(1^\lambda, \sset')$,
    it holds that $|\Pr[\adv(\pp) = 1] - \Pr[\adv(\pp') = 1]| \leq \negl(\lambda)$.
\end{enumerate}

Note that while this PoK primitive is similar to the normal cryptographic PoK primitive,
the primitive we need is weaker:
we just require the attacker to show that he has a subset of a given set of shares.
Thus, we can construct our PoK primitive in very efficient ways.
}
\proofofknowledgedef

\parhead{PoK construction using hashes}
\newcommand{\pokfirstconst}{

We formalize the $\pok$ protocol in \cref{algo:pok}. The hash of the key shares are stored in the smart contract. The submitted key shares are hashed and compared with the hash values stored.
\begin{algorithm}
    \caption{$\pok$}
    \label{algo:pok}
    \begin{algorithmic}[1]
    \Function{$\pok.\Setup$}{$1^\lambda, \sset$}
    \State Set a hash function $h': \{0,1\}^* \to \{0,1\}^\lambda$ modeled as a random oracle
    \State \Return $\pp \gets (1^\lambda, h', (h(s))_{s \in \sset})$
    \EndFunction
    \Function{$\pok.\ProofGen$}{$\pp, \sset'$} 
    \State \Return $(\pi = \sset', k = |\sset'|)$
    \EndFunction
    \Function{$\pok.\verify$}{$\pp, \pi, k$} 
    \State $b \gets 1$
    \State For all $s \in \sset'$, if $h'(s) \not\in \pp$, $b \gets 0$
    \State If $|\pi| \neq k$, $b \gets 0$
    \State \Return $b$
    \EndFunction
\end{algorithmic}    
\label{alg:pok1}
\end{algorithm}
}
\pokfirstconst

\parhead{PoK construction using TEE}
In order to reduce the overhead of the smart contract, we can also let the TEE check the validity of the turned-in key shares.
The attacker encrypts the key shares and send it to the bounty smart contract.
The TEE will provide a signed message confirming the validity of the key shares, using its attestation key, to the smart contract bounty within a specified time.
If the TEE does not respond in time, 
the attacker get automatically paid from the bounty.
If the TEE respond with a signed message saying that the submitted shares are not the correct or the service provider detected that there exists unauthorized signature and get a proof of unauthorized signature from the TEEs, 
the attacker gets nothing.
\newcommand{\poksecondconst}{

}
\poksecondconst

The correctness of this PoK scheme is guaranteed by the correctness of TEE and the correctness of SC.
The extractability is straightforward: the extractor is simply the TEE (which implies the existence requirement in the PoK extractability definition).
The security is guaranteed by the semantic security of the underlying PKE scheme and the trust assumption in TEE \cite{7961949}.  %

\subsection{Bounty workflow}
\label{apd:bountypcode}
When an attacker obtains a set of key shares $\sset'$,
he uses $\sset'$ to generate a proof $(\pi, k)$
to prove that he has already obtained $k$ secret shares.
Upon receiving $(\pi, k)$, the smart contract $\scbounty$ checks the proof to see whether indeed the attacker has the $k$ shares as claimed; if so, it pays the attacker a reward (the amount is determined by the reward function specified in \cref{subsection: reward function design})
and immediately invalidates all the current secret shares (so that it is not possible for the attacker to sell the shares in the market after submitting to the bounty).
Otherwise, the attacker gets nothing.
We formalize the bounty claim workflow in \cref{algo:bountyabstract}.

\begin{algorithm}
\caption{Reward-based Bounty Design}
\label{algo:bountyabstract}
\begin{algorithmic}[1]
    \State Public parameters:
    \begin{enumerate}
        \item $\keycap, T, C(k), K(t), V(k)$ as in \cref{sec:rwdfunccond}
        \item $R(k, t): [1, \keycap] \times [0, T] \to \RR^+$ satisfying the three conditions in \cref{sec:rwdfunccond}.
    \end{enumerate}
    \Function{service provider}{$1^\lambda, \sgk_1, \dots, \sgk_n, m, \keycap$} 
    \State $\sset := (\sgk_1, \dots, \sgk_n)$ \label{step:1inbounty}
    \State Store $\sset$ separately and delete $s$
    \State $\pp \gets \pok.\Setup(1^\lambda, \sset)$
    \State $C \gets \max(R(k, t)), \forall k \in [1, \keycap], t\in[0,T]$
    \State Initiate an SC $\SC(\pp)$ and $C$ units of deposit \label{step:lastinbounty}
    \State Every $T$ units of time: 
    \Indent
        \State pick $m$ shares in $\sset$ as $\sset_m$ and $s \gets \recover(\sset_m)$
        \State terminate the current SC
        \State repeat step \ref{step:1inbounty} to \ref{step:lastinbounty}
    \EndIndent
    \EndFunction
    \Function{Bounty}{$\pp$}  \Comment{Bounty smart contract}
    \If{redFlag == 1}
    \State terminate
    \EndIf
    \State Upon receiving $(\pi, k, \dest)$ at time $t$ \\
    \Comment{$\dest$ is the address where the attacker wants to receive the reward}
    \State If the attacker has already extracted value from the secret, \Return ``Invalid proof''. \label{lin:checkinvalidation}
    \If{$\pok.\verify(\pp, \pi, k) = 1$} 
    \State redFlag $\leftarrow 1$  \Comment{Raise the red flag}
    \State Invalidate the current secret shares \label{lin:invalidation}
    \State Send $R(k, t)$ units of money to address $\dest$
    \State Send $C - R(k, t)$ back to the service provider  \Comment{$C$ }
    \Else
    \State \Return ``Invalid proof''.
    \EndIf
    \EndFunction
    \Function{Attacker}{$\pp, \sset', \dest$}
    \State $(\pk, k) \gets \pok.\ProofGen(\pp, \sset')$
    \State Send $(\pk, k, \dest)$ to SC \Comment{$\dest$ is attacker's account}\\
    \Comment{This process is combined with commit-and-reveal.}
    \EndFunction
\end{algorithmic}
\end{algorithm}

\section{Profitability Analysis}
\label{appendix:profit_analysis}
We discuss how the service provider can make profits.
Naturally,
all the services from a service provider come with a fee.
In our system,
the cost of the service provider includes the operational costs $\operationcost$ 
(e.g., the costs of operating all the TEEs) per $T$ units of time
and $\claimedvalue$, the amounts paid for insurance and bounty claims.

Suppose that there are $\numusers$ users,
and user $i$'s key (that controls $V_i$ amount of assets) is leaked with an independent probability $p_i$ to outsider attacks.
In the analysis, we assume that these parameters and the bounty reward functions are set correctly (i.e., all the (rational) outsider attacks only submit to the bounty instead of abusing the key \footnote{Note that if not the case, we can replace $R_i(k=1,t=0)$ with $V_i$ in the analysis for the same result,
as a user can at most loss $V_i$ units of money.})
Then, the expected money paid for insurance and bounty claims
is $\mu = E(\claimedvalue) = \sum_{i \in [\numusers]} R_i(k = 1, t = 0) \cdot p_i$,
where $R_i(k,t)$ is the reward function for user $i$ as in \cref{sec:rwdfunccond}.%
\footnote{We only need $k=1$ since a rational attacker claims the bounty when they obtain the first share;
since $dR/dt < 0$, we have $R_i(1, t')$ upper bounded by $R_i(1, 0)$.}
The insurance cost from outsiders should just be $0$ as a rational attacker always submits the bounty instead of making unauthorized transactions.
Then, since the insider attacker is simply the service provider,
any cost from insurance claims due to the insider attacker is zero,
as the service provider has obtained the same amount of money from the attack.

Using Chernoff bound,
$\Pr[\claimedvalue \leq (1 + \log(\lambda)\log\log(\lambda)) \mu] \leq \negl(\lambda)$.\footnote{$\log(\lambda)\log\log(\lambda)$ can be replaced with any $\omega(\log(\lambda))$.}
Thus, if the users totally pay $(1 + \log(\lambda)\log\log(\lambda)) \mu + \operationcost$ units of money per $T$ units of time,
the expected gain of the service provider is $\log(\lambda)\log\log(\lambda) \mu$,
and the service provider has a loss with probability $\negl(\lambda)$.

\parhead{Interest cost}
We require 100\% value of wallet collateral for the insurance. 
The collateral is deposited by the service provider and has an interest cost.
This cost is inevitable since we need 100\% collateral to hold the service provider responsible.
Just like the staking system in blockchain systems \cite{EtherumStaking} where the stake is locked, the interest cost is a necessary cost for the security.
Similarly, for the bounty, we have some deposit determined by the reward function,
which also incurs some interest costs.
All these interests are counted as the operational cost above.

\section{Additional Design for the Wallet}
\label{sec:alternativedesign}

In this section,
we introduce some alternative designs that are orthogonal to all the designs we have above,
but are of their own interests.

\subsection{Rate limit}
Recall that in \cref{sec:bounty}, the signing key has a certain value $\amtcap$.
In the use case of a wallet, each user's wallet has some balance.
Naturally, the money deposited in each wallet is the value $\amtcap$ for that wallet.
However, this also means that the balance for each wallet needs to be public .
While this is acceptable in many applications,
we propose additional techniques to mitigate this information leakage.

Instead of having a single wallet for each user,
the service provider sets a rate limit $c$ for each wallet.
For users depositing $D > c$ units of money,
the service provider separates $D$ units of balance into $\ceil{D/c}$ wallets.
Then, each wallet's value at most $\amtcap = c$.
This can also reduce the loss when an attack happens (since a wallet contains at most $c$ units of money).

\subsection{Key rotation based on number of generated traces} 
Currently, user keys are rotated every $T$ time units, following the standard mobile adversary model~\cite{herzberg1995proactive}. 
However, in the specific context of TEE side-channel attacks, the chance of leaking a secret typically increases with the number of uses of that secret, because each use gives an attacker the opportunity to gather new observations about the secret from its vantage point.  A natural extension to our design would then be one where keys are rotated after a fixed number of uses, versus (or perhaps in addition to) the passage of a fixed amount of time.  In a distributed system of TEEs, each holding a share of one private key, this rotation would presumably need to be driven by the secret share used most frequently.
Alternatively,
the keys can be rotated individually and a key usage of that particular key is deferred until the key rotation is finished.

\subsection{Two Step Transactions}
\label{sec:interwallet}
\parhead{Intermediate wallets} 

To better reduce the chance of getting side channels,
we propose the following wallet design:
the service provider keeps two sets of wallets:
storing wallet set $S$ and distribution wallet set $D$.

All the money of the users is stored in some storing wallet.
One user has one storing wallet holding all of her money.
Thus, $|S|$ is the number of users under this design.
Additionally, the company keeps a set of distribution wallets $D$ where $|D|$ can be $\ll |S|$, to be fixed later.

Then, we use techniques like the smart contract to stipulate that:
for any wallet $s \in S$, money in $s$ can only be transferred into some wallet $d \in D$.
Note that the two sets are disjoint (i.e. $D \cup S = \emptyset$).
Wallets in $D$ are acting as normal wallets, allowing regular transactions in and out.

To make a transaction, a client $c$ simply sends a request $(u, X)$ to send $u$ units of money to wallet $X$ (as normal transaction requests above),
where $X$ can be any wallet, not just wallets in $D$ or $S$.
Then, TEE first finds the storing wallet for that client $s_c \in S$,
and makes a transaction from $s_c$ to one of the distribution wallets $d \in D$ (chosen randomly).
Then, after $\numusers$ arrives $d$, immediately initiate a transaction from $d$ to $X$ (i.e., both transactions are done in the same block).

\thenewparagraph{Analysis}
Now we discuss why this approach is a safer design.
Suppose that the adversary only has the key of one of the storing wallets,
there is no benefit that can be obtained by the adversary,
as money cannot be transferred to their own wallets.

Alternatively, if the adversary only has the key to one of the distribution wallets,
the only attack he can do is try to front-run the transaction from $d$ to the target address $X$.
This is because most of the time, the distribution wallets are empty, and when the wallets are not empty,
there is an immediate transaction to make.

The most powerful attack is then possessing one key from the storing wallet $s$ and from the distribution wallets $d$.
Note that this always makes the attack two times harder as the attacker needs to obtain the keys from two independent wallets.
Furthermore, in this case,
the adversary needs to first transfer money from $s \to d$ and then from $d$ to their own wallets.
However, this means that if the service provider actively monitors the mempool or chain,
it will be able to find that there is an unknown transaction $s \to d$,
and make some action (e.g., shut down all the wallets in $D$) to stop the money being transferred out from $d$.
This can be used to mitigate outsider attackers.

Additionally, if a company needs to mount an attack,
a similar analysis holds.
The users can also actively monitor the chain to see if there is any unknown transaction from their own wallet.

Of course, this also means that all honest transactions also have two steps.
This means that the client needs to wait 2 block time to get their transactions completed.
This means that the transaction fee is doubled and the smart contract for checking whether the first transaction is valid (i.e, from $S$ to $D$) also needs some extra gas.
However, this checking can be done using say a hash table, thus with only a cost of $O(1)$.

\thenewparagraph{Dynamic subset of $D$}
One way to further restrict the attacks is to have a dynamic valid subset of $D$ for each wallet $s$.
We first fix some number $b \ll |D|$.
Then, we use a global unpredictable random beacon $r$, and two hash functions $H, G$, modeled as random oracles.
A transaction $s \to d$ is valid if and only if $(H(s || r) \bmod b) = (G(d) \bmod b)$ and $d \in D$.
In this case, for a certain period of time (i.e., how long the random beacon is updated),
only ${\sim}b$ distribution wallet is available for a particular storing wallet,
and thus further restricts the ability of the attackers.
The condition check is still constant time and requires very little extra storage.

\thenewparagraph{Combining with the rate limit}
This can be further combined with the rate limit. Each wallet in $S$ or $D$ has a rate limit $\amtcap$ and thus the benefit per wallet for the attacker gets even smaller.

\subsection{Authentication}
Authentication is provided by the third-party OAuth providers via OAuth in our solution.
However, other forms of authentication could be used as well.
For example, we could use the video of the user holding a written/typed hash of the transaction information as a token for authorizing the transaction.
Face recognition and liveness detection could be used to verify the validity of the token.

\subsection{Proactively track unauthorized transaction}

In this section, we will describe how we can detect the unauthorized transactions, serving two primary purposes:
First, it is more efficient for the system to promptly raise an alert whenever an unauthorized transaction is detected. 
Second, we need to ensure that no unauthorized transactions occur to the corresponding account linked to the submitted key when rewarding the bounty claimer.
This approach reduces reliance solely on the user to submit an insurance claim.
By implementing these measures, we can strengthen the security of the system and provide a more proactive approach to detecting and addressing unauthorized transactions.

To monitor unauthorized transactions on the blockchain, the TEE service provider feeds each block into the TEE and receives a statement of unauthorized transactions.

The statement of unauthorized transactions can be used to raise a red flag and trigger an alert. 
However, it is important to consider the validity of the signing key shares that were committed to the bug bounty before the red flag was raised. 
These key shares should be deemed valid if no unauthorized transactions have occurred in the account associated with the submitted key share(s) before the secret is revealed.

In order for the bounty to be claimed, there should be no unauthorized transactions on the corresponding account. 
If an unauthorized transaction is detected (with a statement of unauthorized from the TEE) before the block when the submitted secret is released, the claimer will not be eligible to receive the reward.

\section{Another Application: Certificate Authority}
\label{sec:ca}
In this section, we modified \snc and apply it to a new application: certificate authority (CA).

\parhead{Background}
Certificate Authorities (CAs) serve as trusted entities, linking identities to public keys. 
Identity verification by CAs can be manual (e.g., via email) or automated through the Automated Certificate Management Environment (ACME) protocol, which uses HTTP or DNS challenges to confirm the domain owner's identity \cite{letsencrypt}.
To monitor the issuance of certificates, they are published in the transparency logs \cite{laurie2014certificate} \cite{TranspartencyLog}.

A Certificate Authority (CA) verifies domain owners' identities and issues certificates to associate them with public keys. 
Public CA is a key point in the security of the Internet.
In order to link their public key and the domain to build a secure channel with their clients, a domain owner needs to request a certificate from some well-trusted CA.
However, currently, the public CA framework largely depends on the credibility of the CA. 
The current framework does not clearly delineate the consequences faced by CA in the event that an unauthorized certificate is issued.
Most often, browsers just distrust a problematic CA, without seeking further compensation \cite{GoogleWosign, GoogleSymantec, TrusticoDigiCert}.

To make the CA more trustworthy, we modify the authorization process of \snc and apply it to enhance the ACME (Automatic Certificate Management Environment) protocol  .
The ACME protocol issues certificates to domain owners upon request, provided they demonstrate ownership of the domain through either the DNS challenge or HTTP challenge~\cite{acme}.
To apply \snc to CA, we need to make the authorization process accountable, 
which can be achieved by requiring a DNS challenge with DNSSEC.
In the DNS challenge, the domain owner is required to place a specific value in the DNS record under her domain name, signed with her DNS key (a feature provided by DNSSEC).
The signature can be used as the proof of authorization.

\parhead{Details}
For concreteness, we follow the specification of Let's encrypt~\cite{letsencrypt} to show how the \snc fits into the public CA.
We modified the authorization process of \snc to fit in the existing CA framework.
The DNS challenge is used instead of the OAuth.

The CA first generates its signing key pairs inside the TEEs.
To make a certificate request,
a DNS challenge is used to verify the domain owner's identity.
We use DNS instead of HTTP as the signature is supported by the DNSSEC \cite{letsencryptDNSSEC} \footnote{ Although DNSSEC is recommended in the Let's encrypt document \cite{letsencryptDNSSEC},
it has not been widely applied yet \cite{elliott2023sad}.
Since our authorization requires DNSSEC, the proposed solution is only applicable when DNSSEC is widely available.},
making the challenge process accountable.
In the DNS challenge, the domain owner puts a random value sent by the CA along with her public key in a DNS TXT record.
The record is signed by her own DNS key. 
Her DNS key is signed by the DNS key of her parent domain.
The root of the chain of trust of the DNS keys is the DNS root key, which is managed by the Internet Assigned Numbers Authority (IANA).
Once the challenge is completed, the domain is associated with its owner's public key, 
and the domain owner can request certificates for her domain using her signing key corresponding to that public key.
The DNS challenge transcripts and the signed requests from users are recorded in the TEE as proof of authorization.
After the challenge is verified,
the certificate is signed by the CA key (generated inside TEE),
and the signed certificate is published in the transparency log in the CA system.

The insurance and bounty is similar to \snc: when a monitor in the CA system detects an unauthorized certificate, the domain owner has the option to file a claim to the blockchain-based insurance;
a reward bounty is set up for outsider attackers.
The value of the compensation for an unauthorized certificate, along with the value of the signing key, is set by the CA.

Note that smart contracts are only used when the CA misbehaves or the breach happens.
For normal procedures,
CA can issue certificates normally without interacting much with those external tools.
Therefore, integrating smart contracts into our framework has a minimal impact on overall efficiency.

\end{document}